\documentstyle[12pt,epsfig]{article}
\catcode`\^^I=13\def^^I{\ \ \ \ \ \ \ \ }
\newcommand{\be}{\begin{equation}}
\newcommand{\ee}{\end{equation}}
\newcommand{\ba}{\begin{eqnarray}}
\newcommand{\ea}{\end{eqnarray}}
\newcommand{\pd}{\partial}
\newcommand{\Tc}{{\mathcal T}_c}
\newcommand{\Td}{{\mathcal T}_d}
\newcommand{\Th}{{\mathcal T}_h}
\newcommand{\bp}{\mathbf p}
\newcommand{\bP}{\mathbf P}
\newcommand{\shat}{\hat{s}}
\newcommand{\that}{\hat{t}}
\newcommand{\uhat}{\hat{u}}
\newcommand{\sighat}{\hat{\sigma}}
\newcommand{\del}{\partial}
\newcommand{\mM}{\overline{\mathcal M}}
\newcommand{\qbar}{\overline q}
\newcommand{\bbar}{\overline b}
\newcommand{\dbar}{\overline d}
\newcommand{\ubar}{\overline u}
\newcommand{\Fbar}{\overline F}
\newcommand{\pr}{{\it Phys. Rev.}}

\newcommand{\pl}{{\it Phys. Lett.}}
\newcommand{\np}{{\it Nuclear Phys.}}

\begin{document}
\begin{center}
{\Huge Topics in Hadronic Physics}\\
\vspace{.25in}
{\Large Alfred Tang}\\
\vspace{.15in}
{\em Physics Department, Baylor University,}\\
{\em P. O. Box 97316, Waco, TX 76798-7316.}\\
\vspace{.1in}
Emails: atang@alum.mit.edu
\end{center}
\vspace{5mm} \noindent

\begin{center}
{\bf Abstract}
\end{center}
\noindent
This work is a pedagogical introduction to the Lund string fragmentation model
and the Feynman-Field hadron production model.  Derivations of important
formulas are worked out in details whenever possible.  An example is
given to show how to
evaluate invariant meson production cross-sections using the Monte Carlo
integration technique.
\vspace{0.5in}
\newpage

\section{Introduction}
The topics discussed in this work are found in original
sources listed as References~\cite{andersson98,field,nr97}.  The materials
shown here are parts of the appendices in my Ph. D. dissertation~\cite{thesis}.
When I was going through these materials the first time, I found some typos and
occasionally inconsistent notations in the original works.  There were also
some elementary questions that one would normally ask, which are not
explained in the texts.  This paper is essentially a record of my notes and
reflections when I was learning these topics.  The Lund model and the
Feynman-Field model are chosen for this discussion because they form the
theoretical foundation of the latter part of my dissertation when
calculating the invariant hadron production cross
sections in the soft and hard $p_T$ regions respectively.  The core of the
latter part of
my Ph. D. dissertation includes (1) a non-perturbative calculation of the
invariant meson production cross sections in the soft $p_T$ region based
on the Lund model, (2) a perturbative calculation of the invariant production
cross sections in the hard $p_T$ region based on the Feynman-Field model,
and (3) the parameterizations of the invariant pion and kaon
production cross sections in both the soft and hard $p_T$ regions.  The
new results listed above will be published as a separate paper and not
to be discussed in this work.

\section{The Area Law of the Lund Model}\label{allm}
The Lund model is discussed extensively in Reference\cite{andersson98} along
with many other interesting topics. We shall focus only on the Area law in
this section.

\subsection{Kinematics}
The quark-antiquark pair of a meson is massless in the string model. In this
case, the concept of the mass of a meson is associated with the mass of
the string field and not the
quarks.  Massless quarks move at the speed of light.  It is not a surprising
result given the fact that string theory predicts that the
ends of open strings move at the speed of light either by imposing a
Neumann boundary condition for open strings~\cite{polchinski98} or
by solving the classical equations of motion~\cite{hatfield92}.  In the highly
relativistic problems, the light-cone coordinates are the natural choice.  In
the $(1+1)$ case, Lorentz transformations can be written as
\be
\left(\begin{array}{c}E'\\p'\end{array}\right)=
\left(\begin{array}{cc}
\gamma & -\gamma v\\
-\gamma v & \gamma \end{array}\right)
\left(\begin{array}{c}E\\p\end{array}\right)\equiv
\left(\begin{array}{cc}
\cosh y & -\sinh y\\
-\sinh y & \cosh y \end{array}\right)
\left(\begin{array}{c}E\\p\end{array}\right),
\ee
\be
\left(\begin{array}{c}t'\\x'\end{array}\right)=
\left(\begin{array}{cc}
\gamma & -\gamma v\\
-\gamma v & \gamma \end{array}\right)
\left(\begin{array}{c}t\\x\end{array}\right)\equiv
\left(\begin{array}{cc}
\cosh y & -\sinh y\\
-\sinh y & \cosh y \end{array}\right)
\left(\begin{array}{c}t\\x\end{array}\right),
\ee
where $\gamma$ is the Lorentz factor
\be
\gamma={1\over\sqrt{1-v^2}}.
\ee
These equations imply
\ba
\gamma&=&\cosh y_p,\\
\gamma v&=&\sinh y_p,\\
v&=&\tanh y_p.\label{tanhy}
\ea
The energy and momentum of a particle can now be written as
\be
E=\gamma m = m\cosh y_p,
\ee
\be
p=\gamma mv=m\sinh y_p.
\ee
The boosted energy $E_b$ and the boosted 1-momentum $p_b$ are given as
\ba
E_b&=&\gamma\,(E-v\,p)\nonumber\\
&=&m\,(\cosh y_p \cosh y - \sinh y_p \sinh y)\nonumber\\
&=&m\,\cosh(y_p-y),\\
p_b&=&\gamma\,(p-v\,E)\nonumber\\
&=&m\,(\sinh y_p \cosh y - \cosh y_p \sinh y)\nonumber\\
&=&m\,\sinh(y_p-y).
\ea
The relativistic velocity addition formula is simple in light-cone 
coordinates~\cite{andersson98},
\be
v'=\tanh y' = {{v-v_b}\over{1-v_b v}}=\tanh(y-y_b),
\ee
illustrating the additivity of rapidity,
\be
y=y'+y_b.
\ee
This simplication motivates the definition of rapidity
\be
y_p={1\over 2}\ln\left({{1+v}\over{1-v}}\right)
={1\over 2}\ln\left({{E+p}\over{E-p}}\right).
\ee
Momenta along the light-cone can be defined as~\cite{andersson98}
\be
p_+ = E+p = m\,\cosh y_p + m\,\sinh y_p = m\,e^{y_p},
\ee
\be
p_- = E-p = m\,\cosh y_p - m\,\sinh y_p = m\,e^{-y_p}.
\ee
With these definitions, boosts are simplified along the
light-cone~\cite{andersson98},
\be
p'_\pm= m\,e^{\pm(y_p-y)}=p_\pm e^{\mp y}.
\ee
Light-cone coordinates can also be defined in configuration space as
\ba
t&\equiv&{m\over\kappa}\cosh y_q,\\
x&\equiv&{m\over\kappa}\sinh y_q,
\ea
where $\kappa$ is the string constant and is used here to give $t$ and $x$ the
correct dimension.  A new subscript is adopted for $y_q$ in
configuration space to distinguish $y_p$ in momentum space.  
These definitions are consistent with the requirement that $v=dx/dt=\tanh y_q$,
as in Eq.~(\ref{tanhy}).  Similar results
are obtained as in the momentum space case,
\be
x_+ = t+x = {m\over\kappa}\cosh y_q + {m\over\kappa}\sinh y_q = 
{m\over\kappa}\,e^{y_q},
\ee
\be
x_- = t-x = {m\over\kappa}\cosh y_q - {m\over\kappa}\sinh y_q = 
{m\over\kappa}\,e^{-y_q},
\ee
\be
x'_\pm= {m\over\kappa}\,e^{\pm(y_q-y)}=x_\pm e^{\mp y}.
\ee

\subsection{Deriving the Area Law}
\begin{figure}[ht]
\begin{center}
\epsfig{file=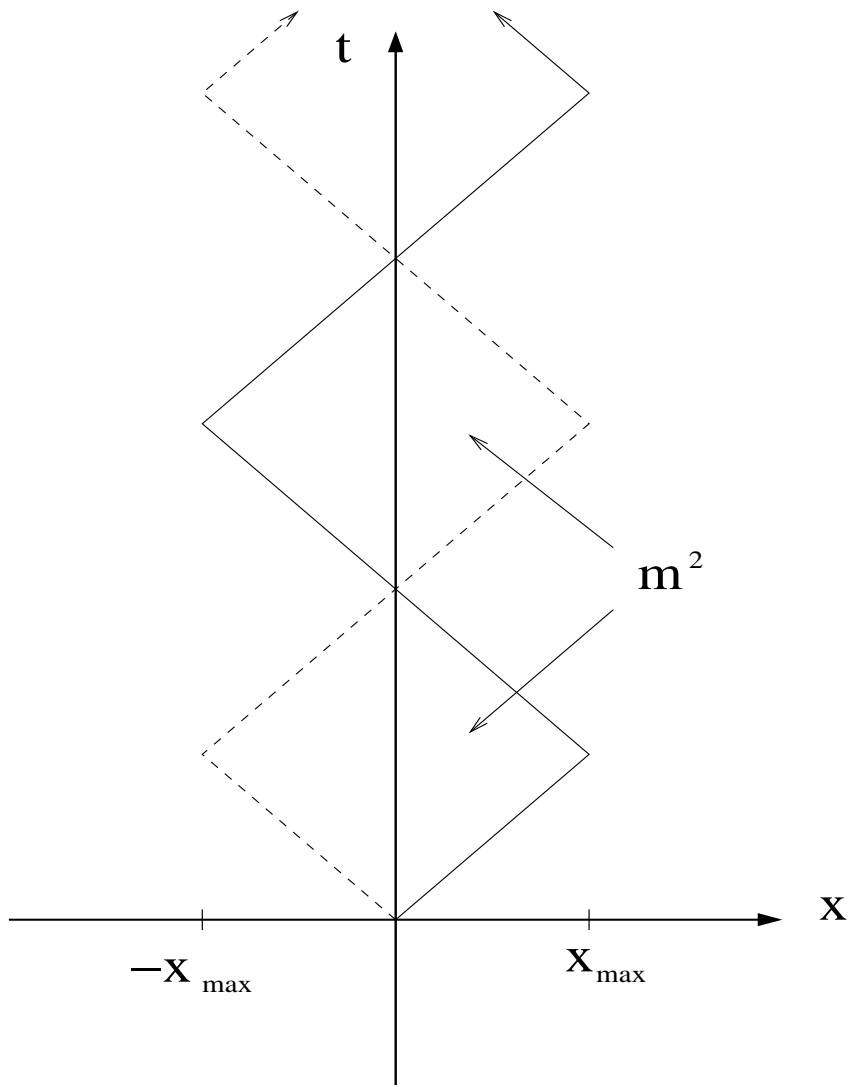,width=12cm,height=15cm}
\caption{\label{yoyo}
The yoyo motion of a quark-antiquark pair confined by a linear potential.  The
two quarks are assumed to be massless.  The mass square of the meson is 
proportional to the area of each square.  The meson shown here is at rest.}
\end{center}
\end{figure}

\begin{figure}[ht]
\begin{center}
\epsfig{file=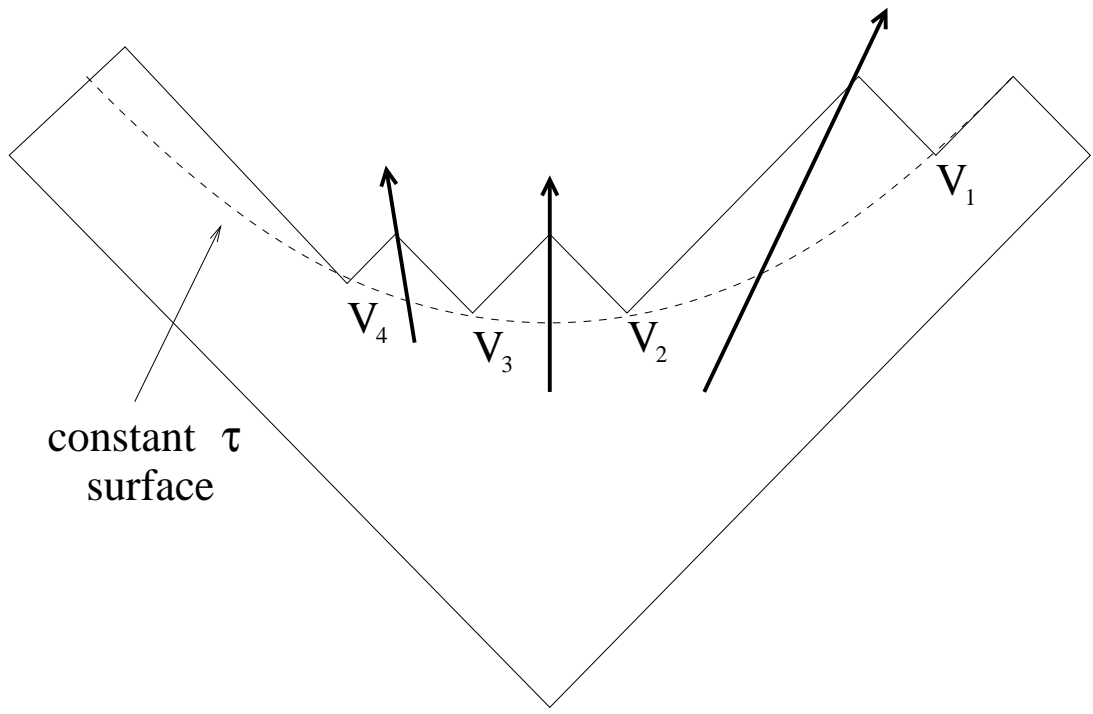,width=15cm,height=10cm}
\caption{\label{breakup}
The breakup of a quark-antiquark pair along a surface of constant proper time
$\tau$.  The bold arrows represents the velocities of the produced mesons.
The breakup points are labelled as vertices $V_1$ to $V_n$.}
\end{center}
\end{figure}

The focus of this paper is primarily mesons.  The force field between a
quark-antiquark pair is presumed to be constant and confined to a flux-tube.
The equation of motion of any one member of the quark-antiquark pair acted
upon by a constant force is~\cite{andersson98}
\be
F={dp\over dt}=-\kappa,
\ee
where $\kappa$ is the string constant.  The solution of the equation
is~\cite{andersson98}
\be
p(t)=p_0-\kappa t=\kappa(t_0-t).\label{yoyo1}
\ee
From $E^2=p^2+m^2$, $E\,dE=p\,dp$ or
\be
{p\over E}={dE\over dp}\label{part1}
\ee
is obtained.  There is also
\be
p=\gamma mv=\gamma m\,{dx\over dt}=E\,{dx\over dt}.\label{part2}
\ee
Together, Eqs. (\ref{part1}) and (\ref{part2}) yield~\cite{andersson98}
\be
{dx\over dt}={p\over E}={dE\over dp},
\ee
such that~\cite{andersson98}
\be
{dE\over dx}={dE\over dp}\,{dp\over dt}\,{dt\over dx}={dp\over dt}=-\kappa.
\ee
Its solution gives the expected QCD flux tube result~\cite{andersson98}
\be
E=\kappa(x_0-x).\label{yoyo2}
\ee
Combining Eqs. (\ref{yoyo1}) and (\ref{yoyo2}) and $m=0$, the relativistic
equation of state is~\cite{andersson98}
\be
m^2=E^2-p^2=\kappa[(x_0-x)^2-(t_0-t)^2]=0.
\ee
Let $x_0=t_0=0$, the equation of motion 
$|x|=t$ is obtained.  The maximum kinetic energy at $x=0$ puts an upper bound
on the displacement of the particle such that $\kappa x_{max}=W$.  The path
of the particle is illustrated in Figure~\ref{yoyo}.  The zig-zag motion of
the particle is the reason for the name ``yoyo state.''  The quark-antiquark
pair of a meson are assumed to be massless.  The mass square of the meson is
proportional to the area of the rectangle defined by the trajectories of the
quark-antiquark pair.  The period $\tau$ 
of the yoyo motion is~\cite{andersson98}
\be
\tau={2E_0\over\kappa},
\ee 
where $E_0$ is the maximum kinetic energy of each quark.  In a boosted
Lorentz frame, the period is transformed as~\cite{andersson98}
\be
\tau'=\tau\cosh y.
\ee
The breakup of a string occurs along a curve of constant proper time such
that the process is Lorentz invariant.  The Lund model assumes that the
produced mesons are ranked, meaning that the production of the $n$-th rank
meson depends on the existence of rank $n-1,\,n-2,\,\dots\,,1$ mesons.
A vertex $V$ is a breakup point and the location $\kappa(x_+,x_-)$ where a
quark-antiquark pair is produced.  The breakup of a string is represented in
Figure~\ref{breakup}.  The produced particles closer to the edges are
the faster moving ones, corresponding to larger rapidities.

\begin{figure}[ht]
\begin{center}
\epsfig{file=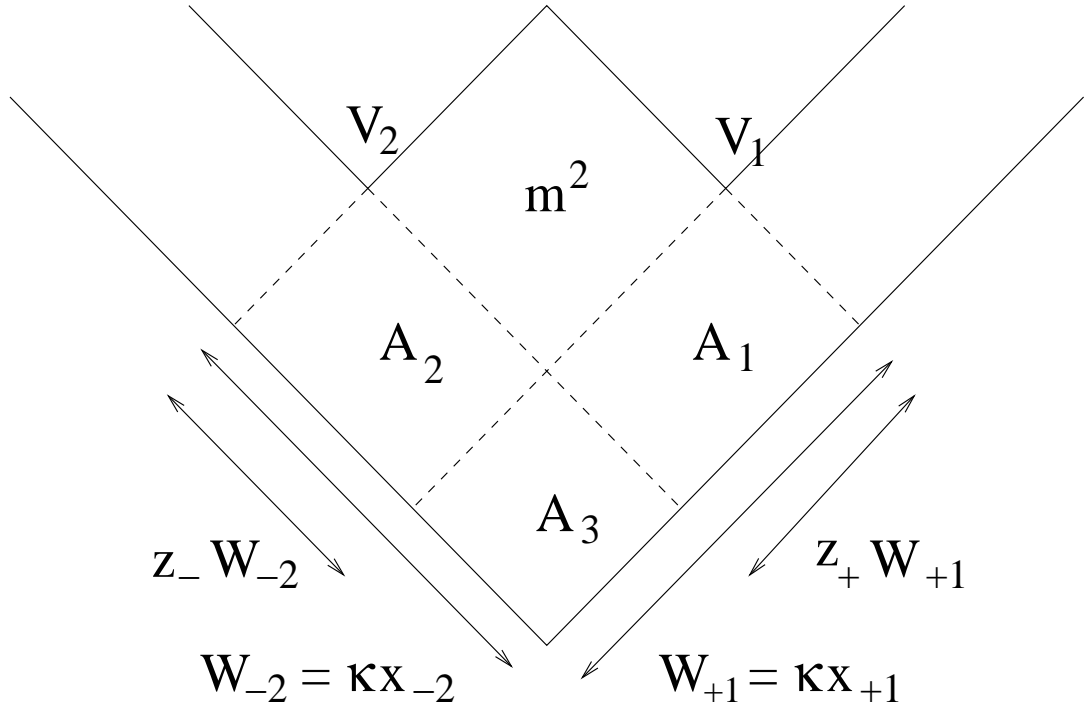,width=15cm,height=10cm}
\caption{\label{g1g2}
The geometry of the kinematics of two adjacent vertices $V_1$ and $V_2$.
$V_1$ and $V_2$ are vertices or spacetime positions which also represent
the energy carried by the string field.
The quark moves along the positive light-cone and the antiquark moves along
the negative light-cone.  The antiquark of $V_1$ combines with the
quark of $V_2$ to produce a meson of mass $m$.  $W_{+1}$ is the energy of
$V_1$ along $x_+$.  $z_+W_{+1}$ is the fraction of energy used to create a
quark from $V_2$.  $W_{-2}$ is the energy of $V_2$ along $x_-$.  $z_-W_{-2}$
is the fraction of energy used to create an antiquark from $V_1$.  $A_1$,
$A_2$, $A_3$ and $m^2$ are the areas of the rectangles.  The figure is
taken from reference~\cite{andersson98}.}
\end{center}
\end{figure}

Let $p_{\pm 0}$ and $p_{\pm j}$ be momentum of the parent
quark and the $j$-th of the $n$ produced quarks moving
along the $x_\pm$ light-cone coordinate respectively.  Then~\cite{andersson98}
\be
p_{\pm 0}=\sum^n_{j=1}p_{\pm j},
\ee
and the momentum fraction at $V_j$ is defined as~\cite{andersson98}
\be
z_{\pm j}={p_{\pm j}\over p_{\pm 0}}.
\ee
In order to simplify notations, $z_j=z_{+j}$ unless specified otherwise.
The invariant interval is
\be
ds^2=dt^2-dx^2=dt^2(1-v^2),
\ee
giving
\be
ds=dt\sqrt{1-v^2}={dt\over\gamma}=d\tau,
\ee
where $\tau$ is the proper time.  The proper time is also
$x_+x_-=(t-x)(t+x)=t^2-x^2=\tau^2$ and can be used as dynamic variable
such that~\cite{andersson98}
\be
\Gamma = \kappa^2 x_+x_-.
\ee
Since $\Gamma$ is proportional to the proper time square $\tau^2$, its Lorentz
invariant property makes it a good kinematic variable in the light-cone
coordinates.
Let $W_{\pm 1}$ and $W_{\pm 2}$ be the kinetic energies along $x_\pm$ at
vertices 1 and 2 respectively.  The following identities can be easily proven
geometrically by calculating the areas of the rectangles $\Gamma_1=A_1+A_3$,
$\Gamma_2=A_2+A_3$ and $m^2$ as shown in Figure~\ref{g1g2}~\cite{andersson98},
\ba
\Gamma_1&=&(1-z_-)W_{-2}W_{+1},\label{g1}\\
\Gamma_2&=&(1-z_+)W_{-2}W_{+1},\label{g2}\\
m^2&=&z_-z_+W_{-2}W_{+1},\label{g3}
\ea
where $m$ is the mass of the produced meson.  Eqs.~(\ref{g1}) and (\ref{g2})
can be expressed with the help of Eq.~(\ref{g3}) as~\cite{andersson98}
\ba
\Gamma_1&=&{m^2(1-z_-)\over z_+z_-},\label{gam1}\\
\Gamma_2&=&{m^2(1-z_+)\over z_+z_-}.\label{gam2}
\ea
Differentiating these equations gives~\cite{andersson98}
\ba
{\pd\Gamma_1\over\pd z_-}=-{m^2\over z_+z_-^2},\label{pg1}\\
{\pd\Gamma_2\over\pd z_+}=-{m^2\over z_+^2z_-},\label{pg2}
\ea
which will be used later in the next section.

\begin{figure}[ht]
\begin{center}
\epsfig{file=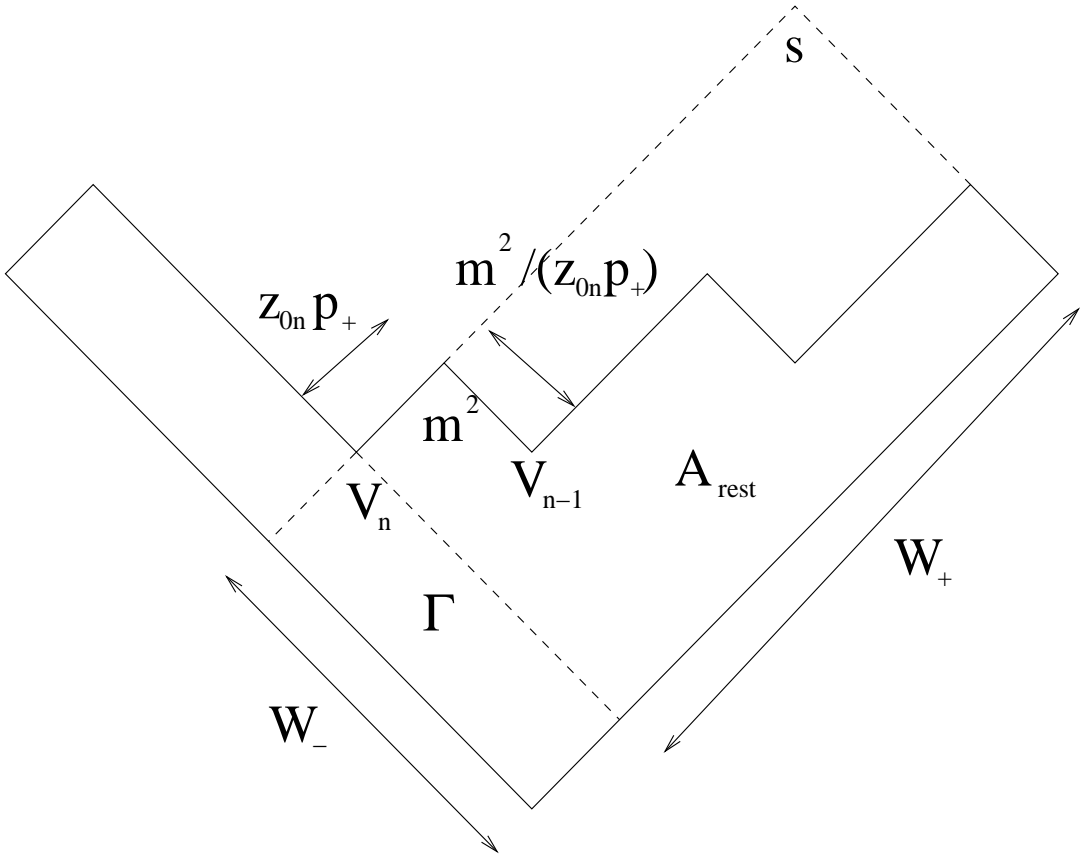,width=15cm,height=12cm}
\caption{\label{geo}
The geometry of the kinematics of involving the total area of the
spacetime diagram such that $A_{total}=\Gamma+A_{rest}$.  The total energy
square $s=W_+W_-$ is represented by the rectangle labelled $s$.  $p_\pm$ is
the energy of the parent quark (antiquark) along the $\pm$ light-cone
coordinates.  $z_{0n}p_+$ is the fraction of energy used to create a
quark from $V_n$.  $m^2/(z_{0n}p_+)$ is the energy used to create an antiquark
from $V_{n-1}$.  The figure is taken from reference~\cite{andersson98} with
minor modifications.}
\end{center}
\end{figure}

Let $H(\Gamma_1)$ be a probability distribution such that
$H(\Gamma_1)\,d\Gamma_1\,dy_1$ is the probability of a quark-antiquark pair 
being produced at the spacetime position $V_1$.  From now on, the symbol $V_n$
also represents the breakup event that leads to the creation of the $n$-th
quark-antiquark pair along a surface of constant $\tau$.  Let
$f(z_+)dz_+$ be the transition probability of
obtaining $V_2$ given that $V_1$ occurs.  The transition probability of 
$V_2$ via $V_1$ is then $H(\Gamma_1)\,d\Gamma_1\,dy_1\,f(z_+)dz_+$.
Similarly the probability of $V_1$ via $V_2$ is
$H(\Gamma_2)\,d\Gamma_2\,dy_2\,f(z_-)dz_-$.  It is reasonable to assume
that the probability of $V_1$ via $V_2$ is equal to that of $V_2$ via
$V_1$ such that~\cite{andersson98}
\be
H(\Gamma_1)\,d\Gamma_1\,dy_1\,f(z_+)dz_+=
H(\Gamma_2)\,d\Gamma_2\,dy_2\,f(z_-)dz_-.\label{prob1}
\ee
The Jacobian $J$ in 
\be
d\Gamma_1\,dz_+=J\,d\Gamma_2\,dz_-,\label{jab1}
\ee
is
\be
J=
\left\Arrowvert \begin{array}{cc}
{\pd\Gamma_1\over\pd\Gamma_2} & {\pd\Gamma_1\over\pd z_-} \\
{\pd z_+\over\pd\Gamma_2} & {\pd z_+\over\pd z_-}
\end{array} \right\Arrowvert
={z_+\over z_-},
\ee
which can be easily computed by using Eqs.~(\ref{pg1}), (\ref{pg2}) and
$\pd\Gamma_1/\pd\Gamma_2=\pd z_+/\pd z_-=0$.
Eq.~(\ref{jab1}) can now be re-expressed as~\cite{andersson98}
\be
d\Gamma_1{dz_+\over z_+}=d\Gamma_2{dz_-\over z_-}.\label{jab2}
\ee
Since rapidity is additive ({\em i. e.} $y_2=y_1+\Delta y$),
\be
dy_1=dy_2.\label{rap}
\ee
With Eqs.~(\ref{jab2}) and (\ref{rap}), Eq.~(\ref{prob1}) is simplified
as~\cite{andersson98}
\be
H(\Gamma_1)z_+f(z_+)=H(\Gamma_2)z_-f(z_-).\label{prob2}
\ee
New definitions are now made to facilitate the solution of the
equation~\cite{andersson98},
\ba
h(\Gamma)&=&\log H(\Gamma),\\
g(z)&=&\log(zf(z)).
\ea
The new definitions transform Eq.~(\ref{prob2}) as~\cite{andersson98}
\be
h(\Gamma_1)+g(z_+)=h(\Gamma_2)+g(z_-).\label{prob3}
\ee
Notice that~\cite{andersson98}
\be
{\pd^2 g(z_+)\over\pd z_+\pd z_-}={\pd^2 g(z_-)\over\pd z_+\pd z_-}=0.
\label{dgdz}
\ee
Differentiate Eq.~(\ref{prob3}) with respect to $z_+$ and $z_-$,
eliminate terms with $\pd^2 g/\pd z_+\pd z_-$ using Eq.~(\ref{dgdz}) and
cancel out factors of $\pd\Gamma/\pd z$ on the both sides of the equation to 
obtain~\cite{andersson98}
\be
{\pd h(\Gamma_1)\over\pd\Gamma_1}+\Gamma_1{\pd^2h(\Gamma_1)\over\pd\Gamma_1^2}=
{\pd h(\Gamma_2)\over\pd\Gamma_2}+\Gamma_2{\pd^2h(\Gamma_2)\over\pd\Gamma_2^2},
\ee
or equivalently~\cite{andersson98}
\be
{d\over d\Gamma}\left(\Gamma\,{dh\over d\Gamma}\right)=-b,
\ee
where $b$ is a constant.  The solution is~\cite{andersson98}
\be
h(\Gamma)=-b\Gamma+a\ln\Gamma+\ln C,
\ee
where $C$ is a constant of integration.  It yields the
distribution~\cite{andersson98}
\be
H(\Gamma)=e^{h(\Gamma)}=C\,\Gamma^a\, e^{-b\Gamma}.\label{dist}
\ee
Substituting Eqs.~(\ref{gam1}) and (\ref{gam2}) into Eq.~(\ref{prob3})
gives~\cite{andersson98}
\ba
&&g_{12}(z_+)+{bm^2\over z_+}+a_1\ln{m^2\over z_+}-a_2\ln{1-z_+\over z_+}
+\ln C_1\nonumber\\
&&\qquad=
g_{21}(z_-)+{bm^2\over z_-}+a_2\ln{m^2\over z_-}-a_1\ln{1-z_-\over z_-}
+\ln C_2,
\ea
where $g_{12}(z_+)$ is $g(z_+)$ of $V_1$ transitioning to $V_2$ and
$g_{21}(z_-)$ is $g(z_-)$ of $V_2$ transitioning to $V_1$.
If $a=a_1=a_2$, the transition probability distribution is~\cite{andersson98}
\be
f(z_j)=N\,{1\over z_j}\,(1-z_j)^a\,e^{-{bm^2\over z_j}},\label{tran1}
\ee
where $N$ is a constant of integration.
Otherwise, the transition probability from $V_\alpha$ to $V_\beta$
is~\cite{andersson98}
\be
f_{\alpha\beta}(z_j)=N_{\alpha\beta}\,{1\over z_j}\,z_j^{a_\alpha}
\,\left({1-z_j\over z_j}\right)^{a_\beta}e^{-{bm^2\over z_j}},\label{tran2}
\ee
where $N_{\alpha\beta}$ is a constant of integration specific to the
vertices $V_\alpha$ and $V_\beta$.
Let $z_{0j}$ be the $j$-th rank momentum fraction scaled with respect to
$p_{+0}$, then~\cite{andersson98}
\ba
z_{01}&=&z_1,\label{z1}\\
z_{02}&=&z_2(1-z_1).\label{z2}
\ea
The probability of two dependent events is the product of the probabilities of
the two individual events.  The existence of the rank-2 hadron depends on
that of the rank-1 hadron according to the Lund Model so that joint
probability of their mutual existence is the product of the two individual
probabilities of $V_1$ and $V_2$.
With Eqs.~(\ref{tran2}-\ref{z2}), the combined distribution of the 1$st$ and 
2$nd$ rank hadrons is~\cite{andersson98}
\ba
f(z_1)dz_1\,f(z_2)dz_2&=&f(z_{01})dz_{01}\,f\left({z_{02}\over 1-z_{01}}\right)
{dz_{02}\over 1-z_{01}}\nonumber\\
&=&{N\,dz_{01}\over z_{01}}\,{N\,dz_{02}\over z_{02}}(1-z_{01})^a
\left(1-{z_{02}\over 1-z_{01}}\right)^a e^{-{bm^2\over z_1}-{bm^2\over z_2}}
\nonumber\\
&=&{N\,dz_{01}\over z_{01}}\,{N\,dz_{02}\over z_{02}}(1-z_{01}-z_{02})^a\,
e^{-b(A_1+A_2)}.
\ea
The geometical identity $A_j=bm^2/z_j$ is used in the last step.
Generalizing the product of two vertices to that of $n$ vertices,
the differential probability for the production of $n$ particles is easily
seen as~\cite{andersson98}
\be
dP(1,\dots,n)=(1-z)^a\prod^n_{j=1}{N\,dz_{0j}\over z_{0j}}\,e^{-bA_j},
\label{dp1}
\ee
where $z=\sum^n_{j=1}z_{0j}$.  Let $p_{0j}=z_{0j}\,p_{+0}$ and
$d^2p=dp_+\,dp_-$.  Use the identity~\cite{andersson98}
\be
\int dC\,dB\,\delta(BC-D)={dB\over B}\label{delta}
\ee
to obtain~\cite{andersson98}
\be
{dz_{01}\over z_{01}}\,{dz_{02}\over z_{02}}=d^2p_{01}\,d^2p_{02}
\delta^+(p_{01}^2-m^2)\,\delta^+(p_{01}^2-m^2).\label{dzz}
\ee
With Eq.~(\ref{dzz}), Eq.~(\ref{dp1}) can be rewritten as~\cite{andersson98}
\be
dP(p_{01},\dots,p_{0n})
=(1-z)^a\prod^n_{j=1}N\,d^2p_{0j}\,\delta^+(p_{0j}^2-m^2)
\,e^{-bA_j}.
\label{dp3}
\ee
It can be shown from simple geometry in Figure~\ref{geo} that the kinetic
energies of the quarks along the $\pm$ light-cones are~\cite{andersson98}
\ba
W_+&=&z\,p_{0+},\\
W_-&=&\sum^n_{j=1}{m_j^2\over z_{0j}p_{0+}},
\ea
and the total kinetic energy square at $V_n$ is~\cite{andersson98}
\be
s=W_+W_-=\sum^n_{j=1}{m^2z\over z_{0j}}.
\ee
The total differential probability of $n$-particle production
is~\cite{andersson98}
\ba
&&dP(z,s;p_{01},\dots,p_{0n})\nonumber\\
&=&dz\,\delta\left(z-\sum^n_{j=1}z_{0j}\right)\,
ds\,\delta\left(s-\sum^n_{j=1}{m^2z\over z_{0j}}\right)
\,dP(p_{01},\dots,p_{0n})\nonumber\\
&=&{dz\over z}\,\delta\left(1-\sum^n_{j=1}u_j\right)\,
ds\,\delta\left(s-\sum^n_{j=1}{m^2\over u_j}\right)\,dP(p_{01},\dots,p_{0n}),
\label{tpp1}
\ea
where $u_j\equiv p_{0+j}/W_{n+}$.  Use the trick~\cite{andersson98}
\ba
&&\delta\left(1-\sum^n_{j=1}{z_{0j}\over z}\right)\,
\delta\left(s-\sum^n_{j=1}{m^2\over u_j}\right)\nonumber\\
&=&\delta\left(W_{n+}-W_{n+}\sum^n_{j=1}u_j\right)\,
\delta\left(W_{n-}-\sum^n_{j=1}{m^2\over u_jW_{n+}}\right)\nonumber\\
&=&\delta\left(W_{n+}-\sum^n_{j=1}p_{0j+}\right)\,
\delta\left(W_{n-}-\sum^n_{j=1}p_{0j-}\right)\nonumber\\
&\equiv&\delta^2\left( p_{rest}-\sum^n_{j=1}p_{0j}\right),
\ea
where $p_{rest}=(W_{n+},W_{n-})$, to re-organize Eq.~(\ref{tpp1})
as~\cite{andersson98}
\ba
dP(z,s;p_{01},\dots,p_{0n})&=&ds\,{dz\over z}\,(1-z)^a\,e^{-b\Gamma}
\delta\left(p_{rest}-\sum^n_{j=1}p_{0j}\right)\nonumber\\
&&\quad\times\prod^n_{j=1}N\,d^2p_{0j}\,\delta^+(p_{0j}^2-m^2)\,
e^{-bA_{rest}},\label{tpp2}
\ea
with the definition~\cite{andersson98}
\be
A_{total}\equiv\sum^n_{j=1}A_j=\Gamma+A_{rest}.
\ee
The claim is that $A_{total}$ is Lorentz invariant.  It is called the
``Area Law''~\cite{andersson01}.  Eq.~(\ref{tpp2}) can be separated in the
external and internal parts as in~\cite{andersson98}
\ba
dP_{ext}&=&ds\,{dz\over z}\,(1-z)^a\,e^{-b\Gamma},\label{ext}\\
dP_{int}&=&\prod^n_{j=1}N\,d^2p_{0j}\,\delta^+(p_{0j}^2-m^2)\,
\delta\left(p_{rest}-\sum^n_{j=1}p_{0j}\right)\,e^{-bA_{rest}}.\label{int}
\ea
The external part contains kinematic variables $s$, $z$ and $\Gamma$.
The internal part contains dynamic variables $p_{0j}$.  Eqs.~(\ref{ext}) and
(\ref{int}) are the final results.

\section{The Parton Model}\label{parton}

The relevant formulas of the parton model in reference~\cite{field} are derived
below.  Several mistakes found in reference~\cite{field} have been corrected
here.

\subsection{External Cross Section Formulas}
\begin{figure}[ht]
\begin{center}
\epsfig{file=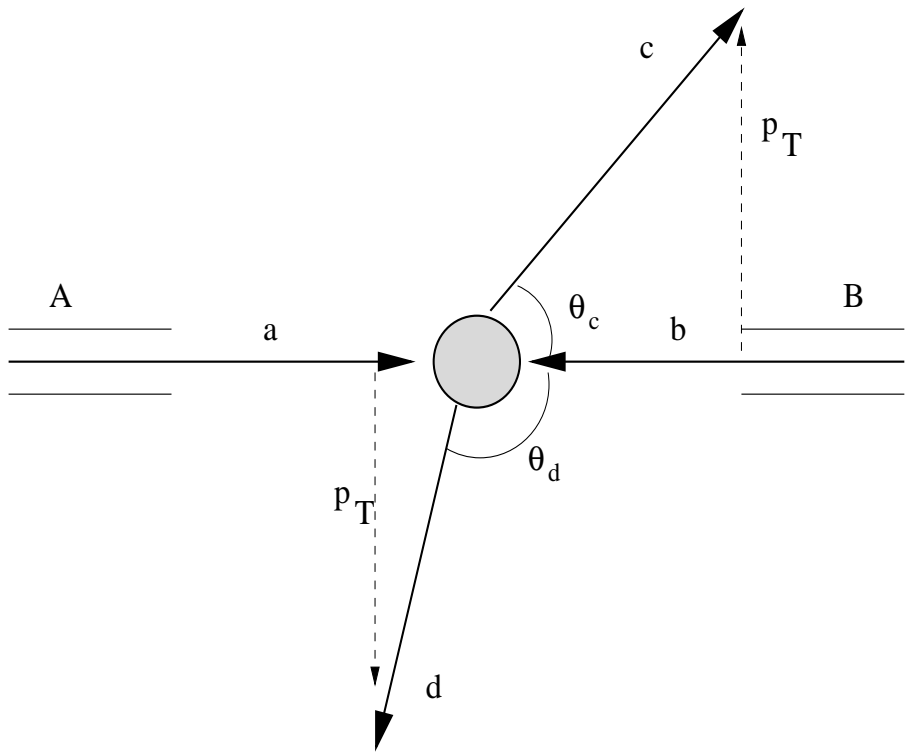,width=12cm,height=10cm}
\caption{\label{partonfig}
Parton $a$ of hadron $A$ collides with parton $b$ of hadron $B$ producing
partons $c$ and $d$.  The transverse momenta, $p_T$, of $c$ and $d$ are equal
and opposite.}
\end{center}
\end{figure}

The Feynman diagram of hadrons $A$ and $B$ scattering into partons
$c$ and $d$, $A+B\to c+d$, is illustrated in Figure~\ref{partonfig}.  The
external Mandelstam variables are defined as
\ba
s&\equiv&(P_A+P_B)^2=2P_A\cdot P_B,\\
t&\equiv&(p_c-P_A)^2=-2p_c\cdot P_A=-{s\over 2}\,x_T\,\Tc,\label{t}\\
u&\equiv&(p_c-P_B)^2=-2p_c\cdot P_B=-{s\over 2}\,x_T\,{1\over\Tc}\label{u},
\ea
where
\be
x_T\equiv{2\,p_T\over\sqrt{s}},
\ee
and
\be
\Tc\equiv\tan{\theta_c\over2}.
\ee
Eq.~(\ref{t}) is proved as follow:
In the high energy limit, $m_i=0,\,\forall i,$ such that
$E_i=|{\mathbf p}_i|$.
\be
E_A=E_B=|{\mathbf P}_A|=|{\bP}_B|={\sqrt{s}\over2}.
\ee
From Figure~\ref{partonfig}, it is obvious that
\be
p_T=|\bp_c|\sin\theta_c.
\ee
\ba
t&=&-2\,p_c\cdot P_A\\
&=&-2\,(E_cE_A-|{\bp}_c||{\bP}_A|\cos\theta_c)\\
&=&-2\,(|{\bp}_c||{\bP}_A|-|{\bp}_c||{\bP}_A|\cos\theta_c)\\
&=&-2\,p_T\,|\bP_A|\,{1-\cos\theta_c\over\sin\theta_c}\\
&=&-2\,\left({x_T\sqrt{s}\over2}\right)\left({\sqrt{s}\over2}\right)
{1-(1-2\sin^2(\theta_c/2))\over2\sin(\theta_c/2)\cos(\theta_c/2)}\\
&=&-{s\over2}\,x_T\,\Tc.
\ea
Similarly,
\ba
u&=&-2\,p_c\cdot P_B\\
&=&-2\,(E_cE_B-|\bp_c||\bP_B|\cos(180^\circ-\theta_c))\\
&=&-2\,(E_cE_B+|\bp_c||\bP_B|\cos\theta_c)\\
&=&-{s\over2}\,x_T\,{1\over\Tc}.
\ea
In the massless limit, the sum rule of external Mandelstam variables is
\be
s+t+u=0.
\ee
A fraction, $x_i$, of the momentum of the incoming hadron $P_i$ makes up
the momentum of parton $p_i$.  More specifically, we have
\ba
p_a&=&x_a\,P_A,\\
p_b&=&x_b\,P_B.
\ea
These imply that
\ba
|\bp_a|&=&{x_a\sqrt{s}\over2},\\
|\bp_b|&=&{x_b\sqrt{s}\over2}.
\ea
The internal Mandelstam variables, $\shat$, $\that$ and $\uhat$, of the
partons can now be written in terms of the external Mandelstam variables,
$s$, $t$ and $u$.
\ba
\shat&=&(p_a+p_b)^2=2p_a\cdot p_b=x_a x_b s,\\
\that&=&(p_c-p_a)^2=-2p_c\cdot p_a=x_a t,\\
\uhat&=&(p_c-p_b)^2=-2p_c\cdot p_b=x_b u.
\ea
The conservation of energy-momentum is given as
\be
p_a+p_b=p_c+p_d.
\ee
From the expressions of the internal and external Mandelstam variables, it
is easily seen that
\be
p_T^2={tu\over s}={\that\uhat\over\shat}.
\ee
In the massless limit, the sum rule of the internal Mendalstam variables is
\be
\shat+\that+\uhat=0,\label{sumhat}
\ee
or equivalently,
\be
x_a x_b s + x_a t + x_b u = 0\label{suma}
\ee
From Eq.~(\ref{suma}), we have
\be
x_a={-x_b u\over x_b s + t}={-x_b u/s\over x_b + t/s}
={x_b x_1\over x_b - x_2},\label{xa}
\ee
where
\ba
x_1&\equiv& -{u\over s} = {1\over 2}\,x_T\,{1\over \Tc},\label{x1}\\
x_2&\equiv& -{t\over s} = {1\over 2}\,x_T\,\Tc.\label{x2}
\ea
Similarly,
\be
x_b={-x_a t\over x_a s + u}={-x_a t/s\over x_a + u/s}
={x_a x_2\over x_a - x_1},\label{xb}
\ee
The expression $\that=(p_c-p_a)^2=(p_d-p_b)^2$ can be considered separately
as
\ba
\that&=&(p_c-p_a)^2=-2p_c\cdot p_a=-{s\over 2}\,x_T\,x_a\,\Tc,\label{t1}\\
\that&=&(p_d-p_b)^2=-2p_d\cdot p_b=-{s\over 2}\,x_T\,x_b\,{1\over\Td},
\label{t2}
\ea
where
\be
\Td\equiv\tan{\theta_d\over2}.
\ee
Eqs.~(\ref{t1}) and (\ref{t2}) imply that
\be
x_a\,\Tc=x_b\,{1\over\Td}.\label{xatcxbtd}
\ee
The internal Mandelstam variables can be re-expressed in terms of the
external angles as follow.
\ba
|\bp_a-\bp_b|&=&|\bp_c|\cos\theta_c + |\bp_d|\cos\theta_d\\
&=&{p_T\over\sin\theta_c}\,\cos\theta_c
+{p_T\over\sin\theta_d}\,\cos\theta_d,
\ea
which implies
\ba
|x_a\,\bP_A-x_b\,\bP_B|&=&\left|x_a\,{\sqrt{s}\over2}-x_b\,{\sqrt{s}\over2}
\right|\\
&=&p_T(\cot\theta_c+\cot\theta_d),
\ea
or equivalently
\ba
x_a-x_b&=&x_T\,(\cot\theta_c+\cot\theta_d)\\
&=&x_T\,\left({1-\tan^2(\theta_c/2)\over2\tan(\theta_c/2)}
+{1-\tan^2(\theta_d/2)\over2\tan(\theta_d/2)}\right)\\
&=&x_T\,\left({1-\Tc^2\over2\Tc}+{1-\Td^2\over2\Td}\right)\\
&=&{1\over2}\,x_T\,\left[\left({1\over\Tc}-\Tc\right)
+\left({1\over\Td}-\Td\right)\right].\label{xaminusxb}
\ea
Eqs.~(\ref{xatcxbtd}) and (\ref{xaminusxb}) together give
\ba
x_a&=&{1\over2}\,x_T\,\left({1\over\Tc}+{1\over\Td}\right),\\
x_b&=&{1\over2}\,x_T\,\left(\Tc+\Td\right).
\ea
Finally, the internal Mendalstam variables can be re-expressed in terms of
the external angles as
\ba
\shat&=&{s\over4}\,x_T^2\,\left(2+{\Tc\over\Td}+{\Td\over\Tc}\right),
\label{shatinangles}\\
\that&=&{s\over4}\,x_T^2\,\left(1+{\Tc\over\Td}\right),\\
\uhat&=&{s\over4}\,x_T^2\,\left(1+{\Td\over\Tc}\right).\label{uhatinangles}
\ea
Eqs.~(\ref{shatinangles})--(\ref{uhatinangles}) will not be used directly in
this work.  However it is interesting to know that the internal Mendalstam
variables can be measured in terms of the external angles.  Another
interesting fact worth mentioning is that
Eqs.~(\ref{xa}) and (\ref{xb}) are still true even if $\Tc$ is replaced by
$\Td$ in Eqs.~(\ref{x1}) and (\ref{x2}).  This statement can be checked easily
by substituting Eq.~(\ref{xatcxbtd}) into Eqs.~(\ref{xa}) and (\ref{xb})
and then solving for $x_a$ and $x_b$.  The interchangability of $\Tc$ and
$\Td$ will become useful later in some of the derivations below.

The differential cross section of parton production for the reaction
$A+B\to c+d$ is
\be
d\sigma=f_{A/a}(x_a,Q^2)dx_a\,f_{B/b}(x_b,Q^2)dx_b\,{d\sighat\over d\that}\,
d\that,\label{parton1}
\ee
where $f_{A/a}(x_a,Q^2)$ and $f_{B/b}(x_b,Q^2)$ are the renormalized parton
distribution functions (PDF's) of parton $a$ in hadron $A$ and of parton $b$ in
hadron $B$ respectively.  Although $Q^2$ has a simple interpretation in QED,
\be
Q^2=(p'-p)^2={1\over2}\,E'E\sin{\theta\over2}
\ee
(where
$p$ and $p'$ are the incoming and outgoing lepton momenta and $\theta$ is the
scattering angle), there is no clear understanding of what is $Q^2$
in hadron collisions.  Some good guesses are
\ba
Q^2&=&4\hat{p}^2_T,\\
Q^2&=&{2\shat\that\uhat\over\shat^2+\that^2+\uhat^2}.
\ea
Although the interest of this work is not parton
production {\it per se}, it serves as an intuitive starting point to introduce
hadron production later.

\begin{figure}[ht]
\begin{center}
\epsfig{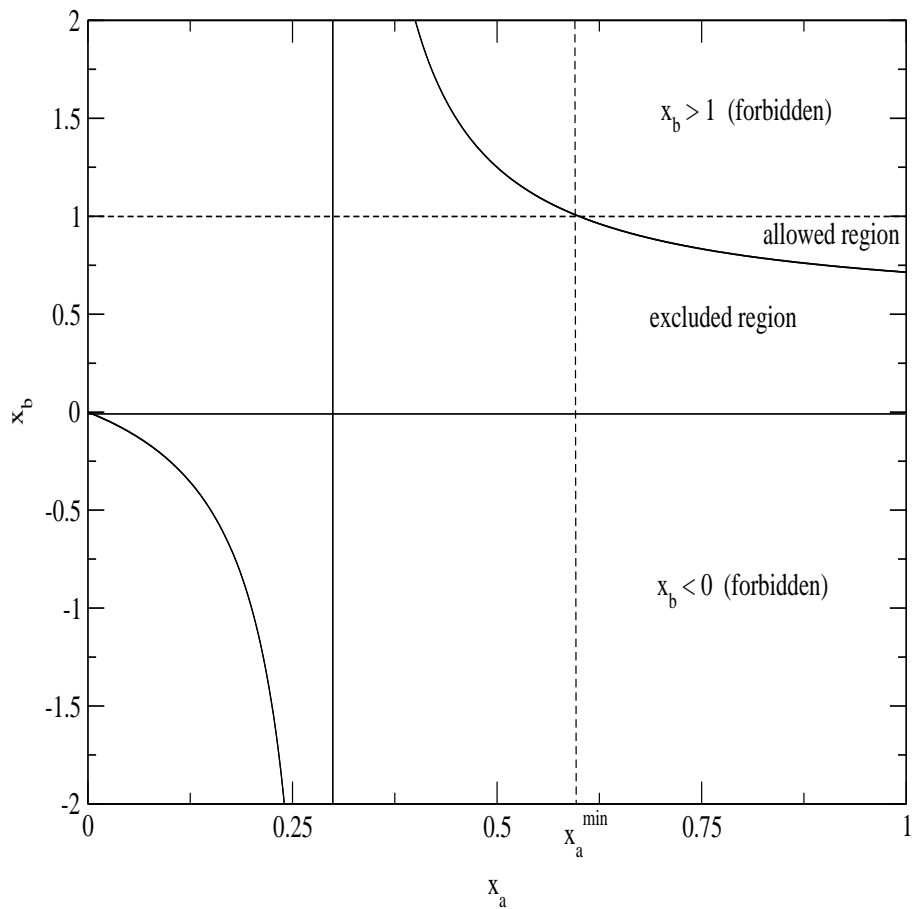}
\caption{\label{xaxbfig}
The plot of $x_a$ versus $x_b$ with $x_b=x_a x_2/(x_a-x_1)$.
$x^{min}_a=x_1/(1-x_2)$.  The values of $x_1=0.5$ and
$x_2=0.3$ are used.}
\end{center}
\end{figure}

The substitution of Eqs.~(\ref{x1}) and (\ref{x2}) into Eq.~(\ref{xa}) gives
\be
x_a x_b-{1\over2}\,x_a x_T\Tc={1\over2}\,x_b x_T\,{1\over\Tc}\label{xaxb}.
\ee
Differentiating $x_b$ in Eq.~(\ref{xaxb}) with respect to $\Tc$ and $x_T$
while keeping $x_a$ fixed gives
\ba
{\del x_b\over\del\Tc}&=&{x_a x_T\Tc - x_b x_T\,\Tc^{-1}\over2x_a\Tc-x_T},
\label{dxdT}\\
{\del x_b\over\del x_T}&=&{x_b + x_a\Tc^2\over2x_a\Tc-x_T}.
\ea
Repeating the same process with $\that=-{s\over2}\,x_T x_a\Tc$ yields
\ba
{\del\that\over\del\Tc}&=&-{s\over2}\,x_T x_a,\\
{\del\that\over\del x_T}&=&-{s\over2}\,x_a\Tc.\label{dtdx}
\ea
The Jacobian now can be computed with Eqs.~(\ref{dxdT})--(\ref{dtdx}) as
\be
{\del(x_b,\that)\over\del(\Tc,x_T)}=
{x_a\,x_b\,x_T\,s\over2x_a\Tc-x_T}={x_a\,x_b\,x_T\,s\over2(x_a-x_1)\Tc}.
\ee
or
\ba
{\del(x_b,\that)\over\del(\theta_c,x_T)}&=&
{\del(x_b,\that)\over\del(\Tc,x_T)}\,{d\Tc\over d\theta_c}\\
&=&{x_a\,x_b\,x_T\,s\over2(x_a-x_1)\Tc}
\left({1-\Tc^2\over2}\right)\\
&=&{s\over2}\,{x_a x_b\over x_a-x_1}\,{x_T\over\sin\theta_c}\label{jacobian1}
\ea
Given that $E=m_T\cosh y$ and $p_z=m_T\sinh y$, $dp_z/dy=E$.  In the
cylindrical coordinates,
\ba
dp^3&=&\pi dp_z\,dp_T^2\\
&=&\pi Edy\,dp_T^2.\label{dp3_1}
\ea
With $p_T^2={s\over2}\,x_T\,dx_T$, 
\be
dp_T^2={s\over2}\,x_T\,dx_T.\label{dpT2}
\ee
And with
\be
y={1\over2}\,\ln{1+\cos\theta_c\over 1-\cos\theta_c},
\ee
there is also
\be
dy={d\theta_c\over\sin\theta_c}.\label{dy}
\ee
Substitute Eqs.~(\ref{dpT2}) and (\ref{dy}) into Eq.~(\ref{dp3_1}) to get
\ba
{dp^3\over E}&=&\pi\,dy\,dp_T^2\\
&=&{\pi\over 2}\,{x_T\,s\over\sin\theta_c}\,d\theta_c\,dx_T.\label{dp3_2}
\ea
Combine the Jacobian in Eq.~(\ref{jacobian1}) with Eq.~(\ref{dp3_2}) and obtain
\be
dx_b\,d\that={1\over\pi}\,{x_a x_b\over x_a-x_1}\,{dp^3\over E}.\label{dxbdt}
\ee
With Eq.~(\ref{dxbdt}), the Lorentz invariant differential parton production
cross section from Eq.~(\ref{parton1}) can be written as
\be
E{d^3\sigma\over dp^3}={1\over\pi}\,f_{A/a}(x_a,Q^2)dx_a\,f_{B/b}(x_b,Q^2)
\,{x_a x_b\over x_a-x_1}\,{d\sighat\over d\that}.\label{parton1_2}
\ee
The inclusive cross section for $A+B\to c+X$ can be obtained from
Eq.~(\ref{parton1_2}) by integrating over $\theta_d$, or equivalently by
integrating over $x_a$ as in $dx_a=(x_a-x_1)\,dy_d$.  The relationship
between $dy_d$ and $dx_a$ can be derived as follow.  Substitute
Eqs.~(\ref{xatcxbtd}), (\ref{x1}) and (\ref{x2}) into Eq.~(\ref{xb}) and
obtain
\be
x_a\Tc\,\Td={x_a\left({1\over2}\,x_T\Td\right)\over x_a-{1\over2}\,x_T\Td}.
\label{xb_2}
\ee
$x_a$ can be isolated from Eq.~(\ref{xb_2}) as
\be
x_a-{1\over2}\,x_T\,{1\over\Td}=x_1.\label{xa_2}
\ee
Differentiating Eq.~(\ref{xa_2}) with $x_T$ and $x_1$ held fixed yields
\ba
dx_a&=&{1\over2}\,x_T\,{1\over\Td^2}\,d\Td\\
&=&{1\over2}\,x_T\,{1\over\Td}\,dy_d\\
&=&{1\over2}\,x_T\,{x_a\Tc\over x_b}\,dy_d\\
&=&{x_a x_2\over x_b}\,dy_d\\
&=&(x_a-x_1)\,dy_d.
\ea
The inclusive cross section for $A+B\to c+X$ is
\be
E{d^3\sigma\over dp^3}={1\over\pi}\int^1_{x_a^{min}}dx_a
\,f_{A/a}(x_a,Q^2)\,f_{B/b}(x_b,Q^2)
\,{x_a x_b\over x_a-x_1}\,{d\sighat\over d\that},\label{parton1_3}
\ee
where
\be
x_a^{min}={x_1\over 1-x_2}={x_T/\Tc\over2-x_T\Tc}.
\ee
The lower limit of the integral in Eq.~(\ref{parton1_3}) is not
zero is that $x_a$, as seen in Eq.~(\ref{xa}).  $x_a<0$ for $x_b<x_2$, which
is not allowed.  At $x_b=x_2$,
$x_a\to\infty$, which is also impossible.  Hence the value of $x_a$ is
limited by a range of $x_b$.  It can be seen in Figure~\ref{xaxbfig} that
$x_a$ is at its minimum when $x_b=1$ and the maximum value of $x_a$ is 1.

Once the formalism of parton production is complete, the derivation of the
inclusive hadron production cross section can be essentially the same.  The
external Mandelstam variables are similarly defined,
\ba
s&\equiv&(P_A+P_B)^2=2P_A\cdot P_B,\\
t&\equiv&(P_h-P_A)^2=-2p_h\cdot P_A=-{s\over2}\,x_T\Th,\\
u&\equiv&(P_h-P_B)^2=-2p_h\cdot P_B=-{s\over2}\,x_T\,{1\over\Th}.
\ea
The internal Mandelstam variables are slightly modified as
\ba
\shat&\equiv&(p_a+p_b)^2=x_a x_b s,\label{shat2}\\
\that&\equiv&(p_c-p_a)^2=-2p_c\cdot p_a=-2{P_h\over z_c}\cdot p_a
={x_a t\over z_c},\\
\uhat&\equiv&(p_c-p_b)^2=-2p_c\cdot p_b=-2{P_h\over z_c}\cdot p_b
={x_b u\over z_c},
\ea
where
\ba
p_a&=&x_a P_A,\\
p_b&=&x_b P_B,\\
P_h&=&z_c p_c.\label{Ph}
\ea
Eq.~(\ref{sumhat}) combined with Eqs.~(\ref{shat2})--(\ref{Ph}) gives
\be
z_c={x_2\over x_b}+{x_1\over x_a},\label{zc}
\ee
with
\ba
x_1&=&-{u\over s}={1\over2}\,x_T\,{1\over\Th},\\
x_2&=&-{t\over s}={1\over2}\,x_T\,\Th,\\
\Th&=&\tan{\theta_{cm}\over2}.
\ea
Differentiate Eq.~(\ref{zc}) to obtain
\ba
{\del z_c\over\del\theta_{cm}}&=&{1\over4}\,\left(-{1\over x_a\Th^2}
+{\Th\over x_b}\right)\,\sec^2{\theta_{cm}\over2},\label{dzdtheta}\\
{\del z_c\over\del x_T}&=&{1\over2}\,\left({1\over x_a\Th}+{\Th\over x_b}\right),\\
{\del\that\over\del\theta}&=&-{1\over4}\,{x_T x_a\over z_c}\,
\sec^2{\theta_{cm}\over2},\\
{\del\that\over\del x_T}&=&-{1\over2}\,{x_a\over z_c}\,\Th.\label{dtdxT_2}
\ea
The Jacobian can be computed from Eqs.~(\ref{dzdtheta})--(\ref{dtdxT_2}),
\be
{\del(z_c,\that)\over\del(\theta_{cm},x_T)}=
{x_T\,s\over2z_c\sin\theta_{cm}}.
\ee
As before,
\be
d\theta_{cm}\,dx_T={2\over\pi s}\,{\sin\theta_{cm}\over x_T}\,{dp^3\over E},
\ee
such that
\be
dz_c\,d\that={1\over\pi z_c}\,{dp^3\over E}.
\ee
The differential hadron production cross section is
\be
d\sigma=f_{A/a}(x_a,Q^2)dx_a\,f_{B/b}(x_b,Q^2)dx_b\,{d\sighat\over d\that}\,
d\that\,D^h_c(z_c,Q^2)dz_c.
\ee
The major difference between the parton and hadron cross section formulas
is the inclusion of the fragmentation function (FF), $D^h_c(z_c,Q^2)$.
The inclusive hadron production cross section is
\be
E\,{d^3\sigma\over dp^3}={1\over\pi}\,\int^1_{x_a^{min}}dx_a
\int^1_{x_b^{min}}dx_b\,f_{A/a}(x_a,Q^2)\,f_{B/b}(x_b,Q^2)\,D^h_c(z_c,Q^2)\,
{1\over z_c}\,{d\sighat\over d\that},
\ee
with
\ba
x_a^{min}&=&{x_1\over1-x_2},\\
x_b^{min}&=&{x_a x_2\over x_a-x_1}.
\ea

\subsection{Internal Cross Section Formulas}
The internal cross sections,
\be
{d\sighat\over d\that}(ab\to cd;\shat,\that)={\pi\alpha_s^2(Q^2)\over\shat^2}\,
|\mM(ab\to cd)|^2,
\ee
are derived by adding the leading order Feynman diagrams~\cite{combridge77}
together, with
\ba
|\mM(q_i q_j\to q_i q_j)|^2&=&
|\mM(q_i \qbar_j\to q_i \qbar_j)|^2={4\over9}\,{\shat^2+\uhat^2\over\that^2},\\
|\mM(q_i q_i\to q_i q_i)|^2&=&{4\over9}\,\left({\shat^2+\uhat^2\over\that^2}
+{\shat^2+\that^2\over\uhat^2}\right)-{8\over27}\,{\shat^2\over\uhat\that},\\
|\mM(q_i \qbar_i\to q_i \qbar_i)|^2&=&{4\over9}\,
\left({\shat^2+\uhat^2\over\that^2}+{\that^2+\uhat^2\over\shat^2}\right)-
{8\over27}\,{\uhat^2\over\shat\that},\\
|\mM(q_i \qbar_i\to gg)|^2&=&{32\over27}\,\left({\uhat^2+\that^2\over
\uhat\that}\right)-{8\over3}\,\left({\uhat^2+\that^2\over\shat^2}\right),\\
|\mM(gg\to q_i \qbar_i)|^2&=&{1\over6}\,\left({\uhat^2+\that^2\over
\uhat\that}\right)-{3\over8}\,\left({\uhat^2+\that^2\over\shat^2}\right),\\
|\mM(q_i g\to q_i g)|^2&=&-{4\over9}\,\left({\uhat^2+\shat^2\over\uhat\shat}
\right)+\left({\uhat^2+\shat^2\over\that^2}\right),\\
|\mM(gg\to gg)|^2&=&{9\over2}\,\left(3-{\uhat\that\over\shat^2}-
{\uhat\shat\over\that^2}-{\shat\that\over\uhat^2}\right),
\ea
and
\be
\alpha_{LO}(Q^2)={12\pi\over(33-2n_f)\log\left({Q^2\over\Lambda^2}\right)},
\ee
\be
\alpha_s(Q^2)=\alpha_{LO}(Q^2)\left[1-{1\over4\pi}\,
{306-38n_f\over33-2n_f}\,\alpha_{LO}(Q^2)\log\log\left({Q^2\over\Lambda^2}
\right)\right],
\ee
where $n_f$ is the number of flavors and $\Lambda_{\overline{MS}}\approx
0.2\,{\rm GeV}^2$ by convention.

\subsection{Fragmentation Functions}
The fragmentation function, $D^h_q(z,Q^2)$, registers the number of hadrons of
type $h$ with energy fraction,
\be
z={2E_h\over Q},
\ee
per $dz$ generated by the initial quark $q$.  Energy conservation gives
\be
\sum_h\int^1_0 zD^h_q(z,Q^2)\,dz=1.
\ee
It is assumed that scaling occurs, {\it i. e.}
\be
D^h_q(z,Q^2)=D^h_q(z).
\ee
The multiplicity of $h$ emerging from $q$ is
\be
\int^1_{z_{min}} D^h_q(z)\,dz,
\ee
where $z_{min}=2m_h/Q$.  The fragmentation function cannot be calculated
exactly by QCD.  Feynman and Field have a semi-analytical
parameterization~\cite{field}, which is now being described below.  Let
$f(\eta)\,d\eta$ be the probability that the first hierarchy (rank 1) meson
leaves fractional momentum $\eta$ to the remaining cascade.  The function
$f(\eta)$ is normalized,
\be
\int^1_0 f(\eta)\,d\eta=1.
\ee
Secondly let $F(z)\,dz$ be the probability of finding a meson (independent
of rank) with fractional momentum $z$ within $dz$ in a jet.  Feynman
and Field assumed a simple stochastic ansatz that probability distribution
is recursive.  This way $F(z)$ satisfies the integral equation
\be
F(z)=f(1-z)+\int^1_z {d\eta\over\eta}\,f(\eta)\,F(z/\eta).\label{Fz}
\ee
The first term gives the probability that the produced meson is rank 1.  The
second term calculates the probability of the production of a higher rank
meson recursively.  Data of deep inelastic scattering experiments suggest
the simple form
\be
zF(z)=f(1-z)=(n+1)(1-z)^n.\label{bogus}
\ee
The power $n=2$ gives a qualitative description of experimental data.

Next the flavor dependent part of the fragmentation function is considered.
Let $\beta_i$ be the probability of the
$q_i \qbar_i$ pair.  The normalization condition imposes that
\be
\sum^{n_f}_{i=1} \beta_i=1.
\ee
The isospin symmetry implies that
\be
\beta_u=\beta_d=\beta.
\ee
Furthermore, data indicate that $\beta_s\approx{1\over2}\,\beta_u$ and
that $\beta_c$ and $\beta_b$ are small.  For a quark of flavor $q$, the
mean number of meson states of flavor $a\bbar$ at $z$ is, in analogy of
Eq.~(\ref{Fz}), given by
\be
P^{a\bbar}_q(z)=\delta_{aq}\beta_b f(1-z)+\int^1_z {d\eta\over\eta}\,f(\eta)\,
\beta_q\,P^{a\bbar}_q(z/\eta).\label{P1}
\ee
The mean number of meson states averaged over all quarks is
\be
P^{a\bbar}_{<q>}(z)=\sum_c \beta_c P^{a\bbar}_c(z),
\ee
so that
\be
P^{a\bbar}_{<q>}(z)=\beta_a\beta_b f(1-z)+\int^1_z {d\eta\over\eta}\,f(\eta)\,
P^{a\bbar}_{<q>}(z/\eta).\label{P2}
\ee
Comparing with Eq.~(\ref{Fz}) yields
\be
P^{a\bbar}_{<q>}(z)=\beta_a\beta_b F(z).\label{P3}
\ee
Comparing Eqs.~(\ref{P1}), (\ref{P2}) and (\ref{P3}) yields
\be
P^{a\bbar}_q(z)=\delta_{aq}\beta_b f(1-z)+\beta_a\beta_b \Fbar(z),\label{P4}
\ee
where
\be
\Fbar(z)=F(z)-f(1-z).\label{Fbar}
\ee
Eqs.~(\ref{bogus}) and (\ref{Fbar}) together give
\be
z\Fbar(z)=(n+1)(1-z)^{n+1}.
\ee
The mean number of meson state, $P^{a\bbar}_q(z)$, is related to the
fragmentation function as in
\be
D^h_q(z)=\sum_{a,b}\Gamma^h_{a\bbar} P^{a\bbar}_q(z),\label{Dhq}
\ee
where $\Gamma^h_{a\bbar}$ is the probability of $h$ containing $a\bbar$.
For example, $\Gamma^{\pi^+}_{u\dbar}=1$ and $\Gamma^{\pi^0}_{u\ubar}=
\Gamma^{\pi^0}_{d\dbar}={1\over2}$.  Substitute Eq.~(\ref{P4}) into
Eq.~(\ref{Dhq}) to obtain
\be
D^h_q(z)=A^h_q f(1-z)+B^h\Fbar(z),
\ee
where
\ba
A^h_q&=&\sum_b\Gamma^h_{q\bbar}\beta_b,\\
B^h&=&\sum_q\beta_q A^h_q.
\ea

As a example, given that $\beta_u=\beta_d=2\beta_s=\beta$ and
$\beta_c=\beta_b=\beta_t=0$, it is easily shown that $\beta=0.4$, and
\ba
A^{\pi^0}_u&=&{1\over2}\,\beta,\\
A^{\pi^0}_d&=&{1\over2}\,\beta,\\
A^{\pi^0}_s&=&0,\\
B^{\pi^0}&=&\beta^2.
\ea

\subsection{Implementing the Feynman-Field Model}\label{ffm}
Feynman and Field calculated the invariant production cross section formula
by incorporating the structure functions (or pardon
distribution functions), $f_{A/a}(x_a, Q^2)$, obtained from DIS experiments
and the fragmentation functions, $D^h_q(z, Q^2)$, derived from a combination
of stochastic arguments and parameterizations of data.  The cross section
formula
\be
E\,{d^3\sigma\over dp^3}={1\over\pi}\sum_{a,b}\int^1_{x_a^{min}}dx_a
\int^1_{x_b^{min}}dx_b\,f_{A/a}(x_a,Q^2)\,f_{B/b}(x_b,Q^2)\,D^h_c(z_c,Q^2)\,
{1\over z_c}\,{d\sighat\over d\that},\label{csf}
\ee
with
\ba
x_a^{min}&=&{x_1\over1-x_2},\\
x_b^{min}&=&{x_a x_2\over x_a-x_1},
\ea
is derived explicitly in Appendix~\ref{parton}.  The invariant cross section
is of interest because it can be used to calculate a number of physical
quantities.  For instance, the particle number ${\mathcal N}_h$ of hadron $h$
can be calculated as~\cite{abramov80}
\be
{\mathcal N}_h={{\mathcal N}_p(1+K_h)\over\sigma^{pp}_{in}}\,\int V(r)\,E\,
{d^3\sigma\over dp^3}\,W(p,r)\,{d^3p\over E}\,d^3r,
\ee
where ${\mathcal N}_p$ is the protons incident on the target; $W(p,r)$ is
the probability of recording a pion produced in the target with coordinate
$r$ and momentum $p$, $V(r)\,d^3r$ is the probability for the incident
proton to interact in the element $d^3r$ in the target; $K_h$ is a
coefficient which takes into account the probability of pion production
as a result of two or more interactions in the target.  Blattnig
{\em et al.}~\cite{steve} calculated the spectral distribution and the
total cross section from the invariant cross section as in
\be
{d\sigma\over dE}=2\pi p\int^{\theta_{max}}_0 d\theta\,E\,
{d^3\sigma\over dp^3}\,\sin\theta,
\ee
and
\be
\sigma=2\pi\int^{\theta_{max}}_0 d\theta\int^{p_{max}}_{p_{min}}\,
dp\,E\,{d^3\sigma\over dp^3}\,{p^2\sin\theta\over\sqrt{p^2+m^2_h}},
\ee
where $m_h$ is the mass of the produced hadron.  These are just some
examples to show why $E\,d^3\sigma/dp^3$ is an interesting quantity
to calculate.

Monte Carlo integration package \texttt{VEGAS} is used to calculate
Eq.~(\ref{csf}).
\texttt{VEGAS} uses a mixed strategy combining importance and stratified
sampling.  The details of this method can be found in
references~\cite{nr97,lepage80}.  Inputs used in \texttt{VEGAS} are
given as follow.
\ba
ndim&=&2,\\
ncall&=&5000,\\
itmx&=&5,\\
nprn&=&0,\\
xl[1]&=&regn[1]={x_1\over1-x_2},\\
xl[2]&=&regn[2]={x_2\over1-x_1},\\
xu[1]&=&regn[3]=1.0,\\
xu[2]&=&regn[4]=1.0.
\ea
The dimension of the integration is $ndim$.  The number of random
calls in the Monte Carlo integration is $ncall$.  The maximum number of
iterations is $itmx$.  The flag $nprn$ controls the initial conditions of
the grids.  In this case, $nprn=0$ signals a cold start.
The random sampling is confined
to an $n$-dimensional rectangular box.  In Lepage's original paper,
the coordinates of the ``lower left corner'' are labelled as $xl[i]$, and
those of the ``upper right corner'' are $xu[i]$.  For a
2-dimensional integral, $i\in\{1,2\}$.  {\em Numerical Recipes} adapted
Lepage's code~\cite{nr97} and puts $xl[i]$ and $xu[i]$ in an array labelled
$regn[j]$ where $j\in\{1,\cdots,2*ndim\}$.
The integrand of Eq.~(\ref{csf}) is implemented as the function
\texttt{fxn} and is expressed in pseudo-code as
\begin{verbatim}
	if (x_b < x_a*x2/(x_a-x1)) return 0;
	else (sum all the terms in the integrand);
\end{verbatim}
The inclusive cross section of $pp\to\pi^+X$ is chosen to illustrate the 
Monte Carlo integration method in this section.
The most accurate parameterization of the parton distributions of proton is
given by the \texttt{CTEQ6} package~\cite{cteq6}.
The QCD running coupling constant, $\alpha_s(Q^2)$, used in the code
is the renormalized
coupling constant described in Appendix~\ref{parton}.  A typical value of
$\Lambda=0.4\,{\rm GeV}$ is used inside $\alpha_s(Q^2)$.  The internal
scattering cross sections of the reactions $q_i\qbar_i\to gg$ and
$gg\to gg$ are excluded from the integral 
because gluons do not fragment into hadrons.
The fragmentation functions used for this calculation are the
original fragmentation functions of Feynman and Field~\cite{ff78}.
For the $pp\to\pi\,X$ reactions, the fragmentation functions are
\be
D^{\pi^0}_u(z)=D^{\pi^0}_d(z)=\left[{\beta\over2}
+\beta^2\left({1-z\over z}\right)\right]\,(n+1)\,(1-z)^n,
\ee
\be
D^{\pi^0}_s(z)=\beta^2\left({1-z\over z}\right)\,(n+1)\,(1-z)^n,
\ee
\be
D^{\pi^-}_d(z)=D^{\pi^+}_u(z)=\left[\beta
+\beta^2\left({1-z\over z}\right)\right]\,(n+1)\,(1-z)^n,
\ee
\be
D^{\pi^{\pm}}_s(z)=D^{\pi^+}_d(z)=D^{\pi^-}_u(z)=
\beta^2\left({1-z\over z}\right)\,(n+1)\,(1-z)^n,
\ee
and for $pp\to K\,X$ reactions, the fragmentation functions are
\be
D^{K^+}_u(z)=D^{K^0}_d(z)={1\over2}\,\left[\beta
+\beta^2\left({1-z\over z}\right)\right]\,(n+1)\,(1-z)^n,
\ee
\be
D^{K^+}_s(z)=D^{K^0}_s(z)=D^{K^+}_d(z)=D^{K^0}_u(z)=
{\beta^2\over2}\,\left({1-z\over z}\right)\,(n+1)\,(1-z)^n,
\ee
\be
D^{\overline{K}^0}_s(z)=D^{K^-}_s(z)=\left[\beta
+{\beta^2\over2}\,\left({1-z\over z}\right)\right]\,(n+1)\,(1-z)^n,
\ee
\be
D^{K^-}_d(z)=D^{K^-}_u(z)=D^{\overline{K}^0}_d(z)=D^{\overline{K}^0}_u(z)=
{\beta^2\over2}\,\left({1-z\over z}\right)\,(n+1)\,(1-z)^n,
\ee
where $\beta=0.4$.
The distributions of $c$, $b$ and $t$ quarks are sufficiently low that
\be
D^h_c(z)=D^h_b(z)=D^h_t(z)=0,
\ee
for any hadron $h$.  Feynman fixed $n=2$ in his original paper.  In this work,
$n$ is a parameter 
freely adjusted to fit data.
There is a subtlety involved in summing all the
parton contributions over $a$ and $b$ in Eq.~(\ref{csf}) that is related to
the relative probabilitistic nature of the parton distributions.  The best
way to explain it is by the way of an analogy.  Suppose there is a room $A$
occupied by 7 boys and 5 girls.  The ratio of boys to girls in that room
is 7:5, or ${7\over12}:{5\over12}$ in normalized format.  Suppose there is
another room $B$ of 14 boys and 10 girls.  The ratio is also 7:5.  
If a superintendent wants to know the number of children in room $B$,
it is not enough to know the ratio of boys to girls.  He must also be given
a integral multiplicative constant.  In this case, 2 times 7:5 or 24 times
${7\over12}:{5\over12}$ gives the
correct head counts which is 14:10.  A similar situation arises in counting
the number of partons in a hadron.  The parton distributions are normalized to
unity so that they give only the relative distributions of the partons.
The parton distributions give only the ratios of the partons in a hadron but
not their numbers.  In order to sum over $a$ and $b$ partons in Eq.~(\ref{csf})
correctly, one must be provided with an integral multiplicative constant for
each of the hadrons $A$ and $B$.
These multiplicative integral constants cannot be
known {\em a prior\'{i}} but are determined {\em a posterior\'{i}}
by fitting data.  In other words, Eq.~(\ref{csf}) can be modified as
\ba
E\,{d^3\sigma\over dp^3}&=&{N_A\,N_B\over\pi}\,
\sum_{\{a,b\}}\,\int^1_{x_a^{min}}dx_a
\int^1_{x_b^{min}}dx_b\nonumber\\
&&\quad\times f_{A/a}(x_a,Q^2)\,f_{B/b}(x_b,Q^2)\,
D^h_c(z_c,Q^2)\,{1\over z_c}\,{d\sighat\over d\that},\label{csf2}
\ea
where $N_A$ and $N_B$ are the multiplicative constants corresponding
to $f_{A/a}(x_a, Q^2)$
and $f_{B/b}(x_b, Q^2)$ respectively and $\sum_{\{a,b\}}$ is the sum over
the parton {\em types} instead of a sum over the partons {\em per se}.
If $A=B$, the overall multiplicative constant, $N_A\,N_B$, is an integer
square.  If $A\ne B$, the overall constant, $N_A\,N_B$,
is still an integer.  In the case of fitting the pQCD calculations
to experimental data of the $pp\to\pi\,X$ reaction, a factor of 100 is missing
if one simply sums over the parton types.  It implies that the multiplicative
constant for the parton distributions of proton is $N_p=10$.  The cross
sections for $K$ production is approximately half of that of $\pi$
production.  It implies that the multiplicative constant for the
$pp\to K\,X$ reaction is $N_p=7$.  For the purpose parameterizing the shape
of the kaon production cross section, the exact value of the scale is not
required.  Therefore
$N_p$ in the $pp\to KX$ reactions is arbitrarily set to be the same as that of
$pp\to\pi X$ reactions at $N_p=10$.  This choice is adequate because
fits to kaon experimental data are not being pursued due to
a lack of experimental data.

\begin{figure}[ht]
\begin{center}
\epsfig{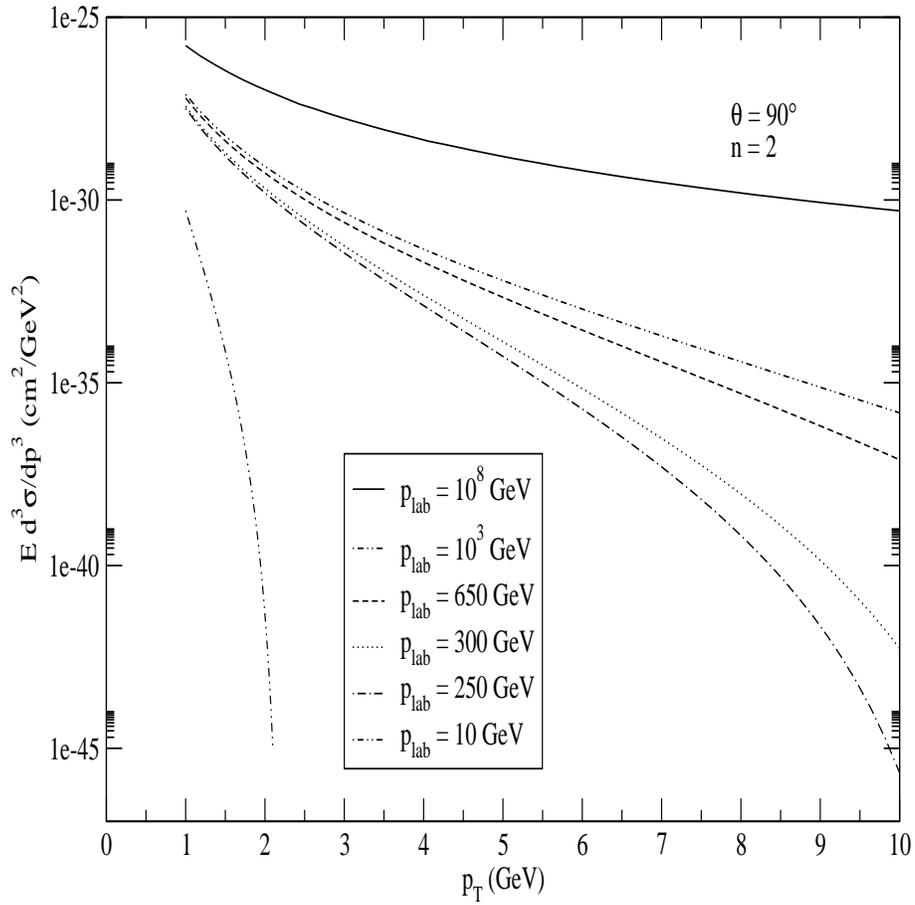}
\caption{\label{test_en}
A sample plot of the Feynman-Field model for $pp\to\pi^+X$
at different energies and $\theta_{cm}=90^{\circ}$.  The plot shows that the
invariant cross section is suppressed at high $p_T$.}
\end{center}
\end{figure}

\begin{figure}[ht]
\begin{center}
\epsfig{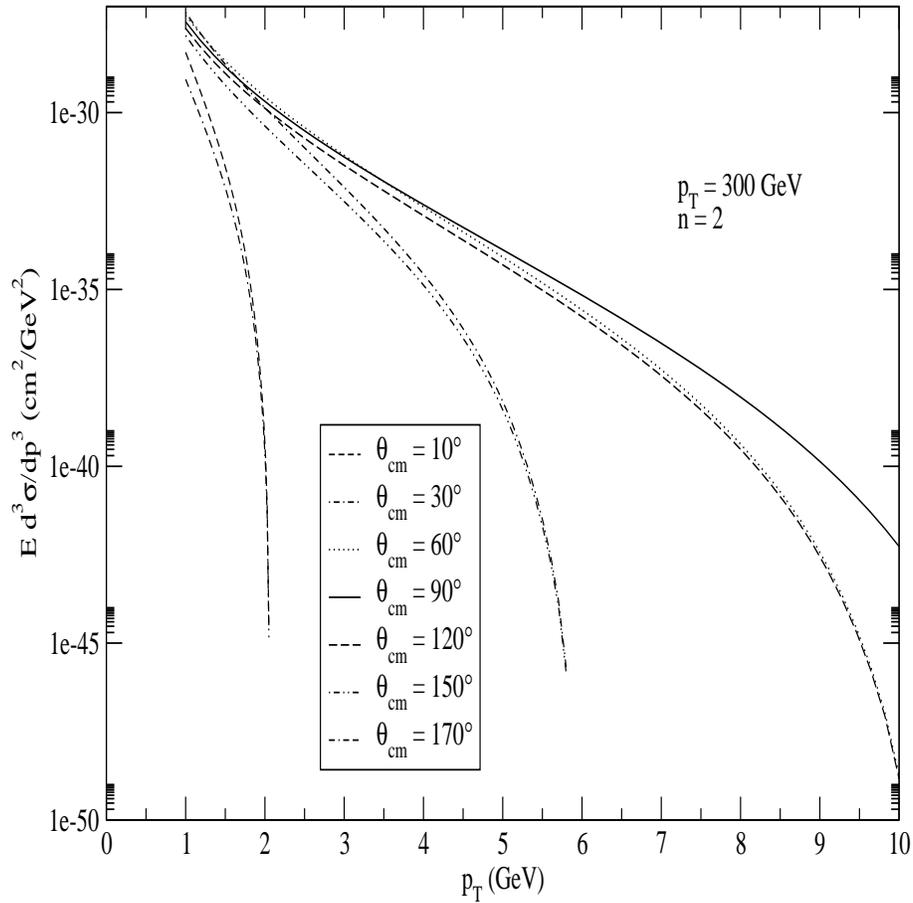}
\caption{\label{test_deg}
A sample plot of the Feynman-Field model for $pp\to\pi^+X$
at different angles.  The plot shows that the
invariant cross section is symmetric around $\theta_{cm}=90^{\circ}$
and is suppressed at $\theta_{cm}\ne90^{\circ}$.}
\end{center}
\end{figure}

\begin{figure}[ht]
\begin{center}
\epsfig{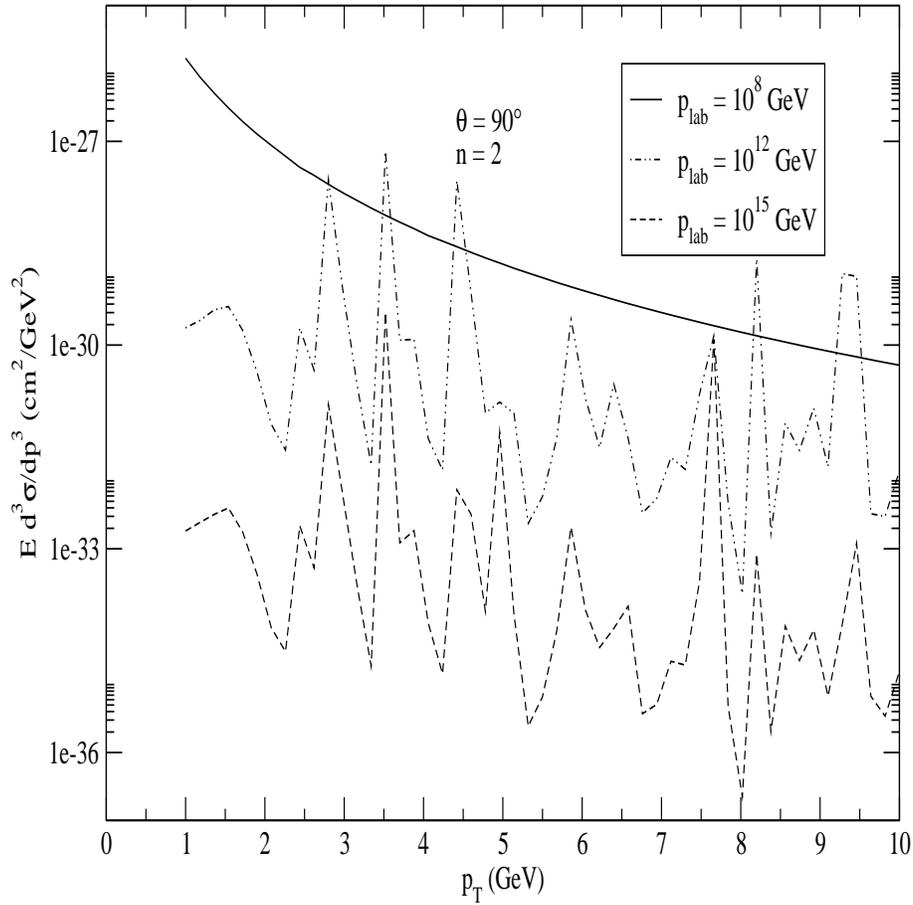}
\caption{\label{test_en2}
A sample plot of the Feynman-Field model for $pp\to\pi^+X$
at excessively high energies and $\theta_{cm}=90^{\circ}$.  The plot shows
that the Monte Carlo code breaks down at ultra-high energies.}
\end{center}
\end{figure}

Figures~\ref{test_en} and \ref{test_deg} illustrate the features of
$E\,d^3\sigma/dp^3$ plotted against $p_T$.  Fig.~\ref{test_en}
shows that the invariant cross section is visibly suppressed at high $p_T$.
Fig.~\ref{test_deg} shows that the suppression increases as $\theta_{cm}$
moves farther away from $90^{\circ}$ and that the suppression is symmetric
around $\theta_{cm}=90^{\circ}$.  Fig.~\ref{test_en2} shows that the
Monte Carlo code is plagued by numerical noise at excessively high
$\sqrt{s}$.  These illustrations summarize the global
properties of the invariant production of cross sections of pions and
kaons.

It is observed that the invariant production cross sections
have the same basic shape regardless of the reactions, {\em i.e.} an
exponentially decaying function of the form $\exp(-\alpha x^\beta)$ at low
$p_T$ and a suppression at high $p_T$ which drops off to zero before the edge
of suppresion at $p_T\le\sqrt{s}/2$.  The cross
section is at its maximum at $p_T=0$ and decreases monotonically in $p_T$.
The Feynman-Field code used in this calculation assumes that
$Q^2=4p_T^2$.  In other words, by combining the previous two statements, the
cross section is at its maximum at $Q^2=0$ and decreases monotonically in
$Q^2$.  This observation indicates that hadron fragmentation
is more favorable at low $Q^2$ in that a parton preserves more kinetic
energy to be made available for hadron fragmentation.  It is also observed
that the cross section is suppressed at high $p_T$ and that the edge of
suppression of the cross section is $\sqrt{s}/2$ at low
$p_T$ and gradually increases toward high $p_T$.  The reason for this
phenomenon is mostly due to the choice of $Q^2=4p_T$ such that
$s\ge Q^2$ or equivalently $\sqrt{s}/2\ge p_T$.  It turns out that the
edge of suppression along $p_T$ is a function of $\sqrt{s}$.  The comments
made so far apply to any angle $\theta_{cm}$ when $\sqrt{s}$ is replaced by
$\sin\theta\sqrt{s}$.  These observations will
be incorporated into the parameterization described in the next section.

\subsection{Monte Carlo Integration}
A sample \texttt{ANSI-C} Monte Carlo program that calculates
$E\,d^3\sigma/dp^3$ written in
using the \texttt{vegas} routine is given below.  The standard
Monte Carlo integration packages are taken from
{\em Numerical Recipes}~\cite{nr97}.
They include \texttt{vegas2.c}, \texttt{rebin.c} and \texttt{ran2.c}.  The
{\em Numeical Recipes} packages need the header files \texttt{nrutil.h},
\texttt{nr.h} and \texttt{nrutil.c} to run.  The \texttt{CTEQ6} package is
downloadable from the web at $\underline{http://www.cteq.org}$.  The
original package is written in \texttt{fortran} but is translated to
\texttt{C} using \texttt{f2c}.  The translated \texttt{CTEQ6} package is
contained in the file \texttt{Ctq6Pdf.c}.  The \texttt{f2c} libraries must be
installed in the system for the translator to work.  When the code is compiled
in unix/linux, the following line must be typed at the command prompt to link
the \texttt{f2c} libraries to the executable:
\begin{verbatim}
    $$ gcc -o pip pip.c -lf2c -lm
\end{verbatim}
where \texttt{pip.c} is the filename of the program.

Below is a sample code that calculates the invariant production cross
section of the $pp\to\pi^+X$ reaction.

\begin{verbatim}
*******************************************************************
#include <stdio.h>
#include <math.h>
#define NRANSI
#include "/recipes/c_d/nrutil.h"
#include "/recipes/c_d/nr.h"
#include "/recipes/c_d/nrutil.c"
#include "/recipes/c_d/vegas2.c"
#include "/recipes/c_d/rebin.c"
#include "/recipes/c_d/ran2.c"
#include "Ctq6Pdf.c"
#define pi 3.1415926535897932384626433
#define lambda2 0.16                            /* part of alpha_s */
#define nf 5.0                      /* number of flavors (topless) */
#define beta 0.4             /* part of the fragmentation function */
#define fac 3.893792923e-28      /* conversion from GeV^-2 to cm^2 */

long idum;

void vegas(double regn[], int ndim, double (*fxn)(double [], double),
	   int init, unsigned long ncall, int itmx, int nprn,
	   double *tgral, double *sd, double *chi2a);
double fxn(double *x, double wt);
doublereal ctq6pdf_(integer *iparton, doublereal *x, doublereal *q);
int setctq6_(integer *iset);
int readtbl_(integer *nu);
integer nextun_(void);
int polint_(doublereal *xa, doublereal *ya, integer *n, 
	    doublereal *x, doublereal *y, doublereal *dy);
doublereal partonx6_(integer *iprtn, doublereal *xx, doublereal *qq);

double x1,x2,s,t,u,Q2,as2;
float plab,d;
integer *ipartona,*ipartonb;
doublereal *q;

MAIN__(void)
{
  int t1,t2,t3,ndim=2,i,imax,init=0,ncall=5000,itmx=5,nprn=0;
  int n=2*ndim;
  integer *iset;
  double pT,dpT,pTmin=1.0,xT,Th;
  float pTmax,tmp;
  double *regn;
  double alo,as,theta;
  double *tgral,*sd,*chi2a;
  FILE *fp;

  regn=dvector(1,n);
  tgral=dvector(1,1);
  sd=dvector(1,1);
  chi2a=dvector(1,1);
  ipartona=lvector(1,1);
  ipartonb=lvector(1,1);
  q=dvector(1,1);

  *iset=1L;                                /* CTEQ6 for MSbar */
  fp=fopen("out.dat","w");

  printf("plab = ");
  scanf("%f",&plab);
  printf("theta = ");
  scanf("%f",&tmp);
  printf("d = ");
  scanf("%f",&d);
  printf("pTmax = ");
  scanf("%f",&pTmax);
  printf("length (10 or 100): ");
  scanf("%d",&imax);

  t1=time(0);

  setctq6_(iset);          /* initialization for CTEQ6 routines */
  dpT=(pTmax-pTmin)/imax;
  pT=pTmin;
  theta=tmp*pi/180.0;
  s=2.0*0.938272*plab;
  Th=tan(theta/2.0);
  regn[3]=1.0;                                        /* xa_max */
  regn[4]=1.0;         /* xb_max */

  t2=time(0);

  for(i=0;i<=imax;i++) {
    Q2=4.0*pT*pT;
    q[1]=sqrt(Q2);
    alo=12.0*pi/(33.0-2.0*nf)/log(Q2/lambda2);
    as=alo*(1.0-(306.0-38.0*nf)/4.0/pi/(33.0-2.0*nf)*alo
	*log(log(Q2/lambda2)));
    as2=as*as;
    xT=2.0*pT/sqrt(s);
    x1=0.5*xT/Th;
    x2=0.5*xT*Th;
    t=-s*x2;
    u=-s*x1;
    regn[1]=x1/(1.0-x2); /* xa_min */
    regn[2]=x2/(1.0-x1); /* xb_min */
    vegas(regn, ndim, fxn, init, ncall, itmx, nprn,
	  &tgral[1], &sd[1], &chi2a[1]);
    fprintf(fp,"%f\t%e\n",pT,fac*tgral[1]*100.0);
    pT+=dpT;
  }

  t3=time(0);

  fclose(fp);
  free_dvector(regn,1,n);
  free_dvector(tgral,1,1);
  free_dvector(sd,1,1);
  free_dvector(chi2a,1,1);
  free_lvector(ipartona,1,1);
  free_lvector(ipartonb,1,1);
  free_dvector(q,1,1);
  printf("\ntime used for overheads = %d seconds\n",t2-t1);
  printf("\ntime used for integrals = %d seconds\n\n\a",t3-t2);
}

double fxn(double *x, double wt) {
  double cs,z,Fbar,f,tmp;
  double sh,th,uh;        /* internal mandelstam vars */
  double sh2,th2,uh2;
  double M1,M2,M3,M4,M5;  /* inv amp */
  double Du,Dd,Ds;        /* fragmentation functions */

  z=x2/x[2]+x1/x[1];
  sh=x[1]*x[2]*s;
  th=x[1]*t/z;
  uh=x[2]*u/z;
  sh2=sh*sh;
  th2=th*th;
  uh2=uh*uh;
  M1=4.0/9.0*(sh2+uh2)/th2;
  M2=4.0/9.0*((sh2+uh2)/th2+(sh2+th2)/uh2)-8.0/27.0*sh2/uh/th;
  M3=4.0/9.0*((sh2+uh2)/th2+(th2+uh2)/sh2)-8.0/27.0*uh2/sh/th;
  M4=(uh2+th2)/uh/th/6.0-3.0/8.0*(uh2+th2)/sh2;
  M5=-4.0/9.0*(uh2+sh2)/uh/sh+(uh2+sh2)/th2;
  f=(d+1.0)*pow(1.0-z,d);
  Fbar=(1.0/z-1.0)*f;
  Ds=Dd=beta*beta*Fbar;
  Du=beta*f+Ds;

  if (x[2]<x[1]*x2/(x[1]-x1)) cs=0.0;
  else {
    cs=0.0;

    /* su->s */
    ipartona[1]=3L;
    ipartonb[1]=1L;
    cs+=ctq6pdf_(&ipartona[1],&x[1],&q[1])
	*ctq6pdf_(&ipartonb[1],&x[2],&q[1])*M1*Ds;

    /* subar->s */
    ipartona[1]=3L;
    ipartonb[1]=-1L;
    cs+=ctq6pdf_(&ipartona[1],&x[1],&q[1])
	*ctq6pdf_(&ipartonb[1],&x[2],&q[1])*M1*Ds;

    /* us->u */
    ipartona[1]=1L;
    ipartonb[1]=3L;
    cs+=ctq6pdf_(&ipartona[1],&x[1],&q[1])
	*ctq6pdf_(&ipartonb[1],&x[2],&q[1])*M1*Du;

    /* usbar->u */
    ipartona[1]=1L;
    ipartonb[1]=-3L;
    cs+=ctq6pdf_(&ipartona[1],&x[1],&q[1])
	*ctq6pdf_(&ipartonb[1],&x[2],&q[1])*M1*Du;

    /* ub->u */
    ipartona[1]=1L;
    ipartonb[1]=5L;
    cs+=ctq6pdf_(&ipartona[1],&x[1],&q[1])
	*ctq6pdf_(&ipartonb[1],&x[2],&q[1])*M1*Du;

    /* ubar->u */
    ipartona[1]=1L;
    ipartonb[1]=-5L;
    cs+=ctq6pdf_(&ipartona[1],&x[1],&q[1])
	*ctq6pdf_(&ipartonb[1],&x[2],&q[1])*M1*Du;

    /* uc->u */
    ipartona[1]=1L;
    ipartonb[1]=4L;
    cs+=ctq6pdf_(&ipartona[1],&x[1],&q[1])
	*ctq6pdf_(&ipartonb[1],&x[2],&q[1])*M1*Du;

    /* ucbar->u */
    ipartona[1]=1L;
    ipartonb[1]=-4L;
    cs+=ctq6pdf_(&ipartona[1],&x[1],&q[1])
	*ctq6pdf_(&ipartonb[1],&x[2],&q[1])*M1*Du;

    /* sd->s */
    ipartona[1]=3L;
    ipartonb[1]=2L;
    cs+=ctq6pdf_(&ipartona[1],&x[1],&q[1])
	*ctq6pdf_(&ipartonb[1],&x[2],&q[1])*M1*Ds;

    /* sdbar->s */
    ipartona[1]=3L;
    ipartonb[1]=-2L;
    cs+=ctq6pdf_(&ipartona[1],&x[1],&q[1])
	*ctq6pdf_(&ipartonb[1],&x[2],&q[1])*M1*Ds;

    /* sb->s */
    ipartona[1]=3L;
    ipartonb[1]=5L;
    cs+=ctq6pdf_(&ipartona[1],&x[1],&q[1])
	*ctq6pdf_(&ipartonb[1],&x[2],&q[1])*M1*Ds;

    /* sbbar->s */
    ipartona[1]=3L;
    ipartonb[1]=-5L;
    cs+=ctq6pdf_(&ipartona[1],&x[1],&q[1])
	*ctq6pdf_(&ipartonb[1],&x[2],&q[1])*M1*Ds;

    /* sc->s */
    ipartona[1]=3L;
    ipartonb[1]=4L;
    cs+=ctq6pdf_(&ipartona[1],&x[1],&q[1])
	*ctq6pdf_(&ipartonb[1],&x[2],&q[1])*M1*Ds;

    /* scbar->s */
    ipartona[1]=3L;
    ipartonb[1]=-4L;
    cs+=ctq6pdf_(&ipartona[1],&x[1],&q[1])
	*ctq6pdf_(&ipartonb[1],&x[2],&q[1])*M1*Ds;

    /* ds->d */
    ipartona[1]=2L;
    ipartonb[1]=3L;
    cs+=ctq6pdf_(&ipartona[1],&x[1],&q[1])
	*ctq6pdf_(&ipartonb[1],&x[2],&q[1])*M1*Dd;

    /* dsbar->d */
    ipartona[1]=2L;
    ipartonb[1]=-3L;
    cs+=ctq6pdf_(&ipartona[1],&x[1],&q[1])
	*ctq6pdf_(&ipartonb[1],&x[2],&q[1])*M1*Dd;

    /* db->d */
    ipartona[1]=2L;
    ipartonb[1]=5L;
    cs+=ctq6pdf_(&ipartona[1],&x[1],&q[1])
	*ctq6pdf_(&ipartonb[1],&x[2],&q[1])*M1*Dd;

    /* dbbar->d */
    ipartona[1]=2L;
    ipartonb[1]=-5L;
    cs+=ctq6pdf_(&ipartona[1],&x[1],&q[1])
	*ctq6pdf_(&ipartonb[1],&x[2],&q[1])*M1*Dd;

    /* dc->d */
    ipartona[1]=2L;
    ipartonb[1]=4L;
    cs+=ctq6pdf_(&ipartona[1],&x[1],&q[1])
	*ctq6pdf_(&ipartonb[1],&x[2],&q[1])*M1*Dd;

    /* dcbar->d */
    ipartona[1]=2L;
    ipartonb[1]=-4L;
    cs+=ctq6pdf_(&ipartona[1],&x[1],&q[1])
	*ctq6pdf_(&ipartonb[1],&x[2],&q[1])*M1*Dd;

    /* ud->u */
    ipartona[1]=1L;
    ipartonb[1]=2L;
    cs+=ctq6pdf_(&ipartona[1],&x[1],&q[1])
	*ctq6pdf_(&ipartonb[1],&x[2],&q[1])*M1*Du;

    /* udbar->u */
    ipartona[1]=1L;
    ipartonb[1]=-2L;
    cs+=ctq6pdf_(&ipartona[1],&x[1],&q[1])
	*ctq6pdf_(&ipartonb[1],&x[2],&q[1])*M1*Du;

    /* du->d */
    ipartona[1]=2L;
    ipartonb[1]=1L;
    cs+=ctq6pdf_(&ipartona[1],&x[1],&q[1])
	*ctq6pdf_(&ipartonb[1],&x[2],&q[1])*M1*Dd;

    /* dubar->d */
    ipartona[1]=2L;
    ipartonb[1]=-1L;
    cs+=ctq6pdf_(&ipartona[1],&x[1],&q[1])
	*ctq6pdf_(&ipartonb[1],&x[2],&q[1])*M1*Dd;

    /* uu->u */
    ipartona[1]=1L;
    ipartonb[1]=1L;
    cs+=ctq6pdf_(&ipartona[1],&x[1],&q[1])
	*ctq6pdf_(&ipartonb[1],&x[2],&q[1])*M2*Du;

    /* dd->d */
    ipartona[1]=2L;
    ipartonb[1]=2L;
    cs+=ctq6pdf_(&ipartona[1],&x[1],&q[1])
	*ctq6pdf_(&ipartonb[1],&x[2],&q[1])*M2*Dd;

    /* ss->s */
    ipartona[1]=3L;
    ipartonb[1]=3L;
    cs+=ctq6pdf_(&ipartona[1],&x[1],&q[1])
	*ctq6pdf_(&ipartonb[1],&x[2],&q[1])*M2*Ds;

    /* uubar->u */
    ipartona[1]=1L;
    ipartonb[1]=-1L;
    cs+=ctq6pdf_(&ipartona[1],&x[1],&q[1])
	*ctq6pdf_(&ipartonb[1],&x[2],&q[1])*M3*Du;

    /* ddbar->d */
    ipartona[1]=2L;
    ipartonb[1]=-2L;
    cs+=ctq6pdf_(&ipartona[1],&x[1],&q[1])
	*ctq6pdf_(&ipartonb[1],&x[2],&q[1])*M3*Dd;

    /* ssbar->s */
    ipartona[1]=3L;
    ipartonb[1]=-3L;
    cs+=ctq6pdf_(&ipartona[1],&x[1],&q[1])
	*ctq6pdf_(&ipartonb[1],&x[2],&q[1])*M3*Ds;

    /* gg->u,d,s */
    ipartona[1]=0L;
    ipartonb[1]=0L;
    cs+=ctq6pdf_(&ipartona[1],&x[1],&q[1])
	*ctq6pdf_(&ipartonb[1],&x[2],&q[1])*M4*(Du+Dd+Ds);

    /* ug->u */
    ipartona[1]=1L;
    ipartonb[1]=0L;
    cs+=ctq6pdf_(&ipartona[1],&x[1],&q[1])
	*ctq6pdf_(&ipartonb[1],&x[2],&q[1])*M5*Du;

    /* dg->d */
    ipartona[1]=2L;
    ipartonb[1]=0L;
    cs+=ctq6pdf_(&ipartona[1],&x[1],&q[1])
	*ctq6pdf_(&ipartonb[1],&x[2],&q[1])*M5*Dd;

    /* sg->s */
    ipartona[1]=3L;
    ipartonb[1]=0L;
    cs+=ctq6pdf_(&ipartona[1],&x[1],&q[1])
	*ctq6pdf_(&ipartonb[1],&x[2],&q[1])*M5*Ds;

    cs*=as2/sh2/z;
  }
  return cs;
}

#undef NRANSI
\end{verbatim}

\section{Conclusion}
The Lund and Feynman-Field models are derived and implemented numerically
in this work.  New results based on these models will published as a
separate paper.

\newpage

\newpage

\newpage

\newpage
\begin{figure}[ht]
\begin{center}
\epsfig{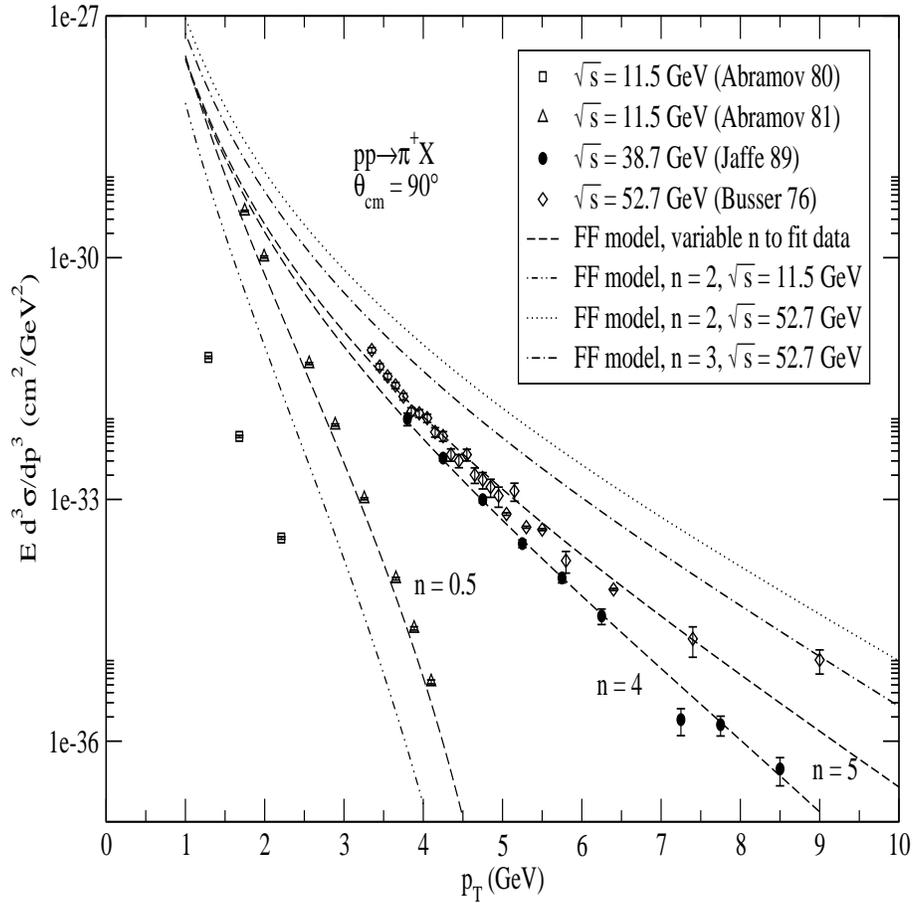}
\caption{\label{pip_demo}
Comparisons of the Feynman-Field model fit for $pp\to\pi^+X$ for various
$n$ at $\theta_{cm}=90^{\circ}$.  The references of the data sets Abromov 80,
Abramov 81, Jaffe 89 and
Busser 76 are \np, {\bf B}173 348 (1980), {\em ZETFP} 33 304 (1981),
\pr, {\bf D}40 2777 (1989) and
\np, {\bf B}106 1 (1976) respectively.}
\end{center}
\end{figure}

\newpage

\newpage

\newpage

\newpage

\newpage

\newpage

\end{document}